\RequirePackage{fix-cm}
\documentclass{svjour3}                     
\smartqed  
\usepackage{graphicx}

\usepackage{graphicx}
\usepackage{xcolor}
\usepackage{hyperref}
\usepackage[numbers]{natbib}
\usepackage{longtable}
\usepackage{paralist}
\usepackage{array}
\usepackage{booktabs}
\usepackage{multirow}
\usepackage{adjustbox}
\newcommand*\rot{\rotatebox{90}}
\usepackage{pdflscape}

\usepackage{tabularx}
\newcolumntype{R}[1]{>{\raggedleft\arraybackslash}p{#1}}

\journalname{Empirical Software Engineering}

\begin{document}

\title{Naming the Pain in Requirements Engineering}
\subtitle{Contemporary Problems, Causes, and Effects in Practice}


\author{D. M\'{e}ndez Fern\'{a}ndez \and
        S. Wagner \and
        M. Kalinowski \and
        M. Felderer \and
        P. Mafra \and
        A. Vetr\`{o} \and  
        T. Conte \and
        M.-T. Christiansson \and
        D. Greer \and
        C. Lassenius \and
        T. M\"annist\"o \and
        M. Nayebi \and
        M. Oivo \and
        B. Penzenstadler \and
        D. Pfahl \and
        R. Prikladnicki \and
        G. Ruhe \and
        A. Schekelmann \and
        S. Sen \and
        R. Spinola \and
        A. Tuzcu \and
        J. L. de la Vara \and
        R. Wieringa
}


\institute{Daniel M\'{e}ndez Fern\'{a}ndez \at
              Technische Universit\"at M\"unchen, Germany\\
              Tel.: +49-89-28917056\\
              \email{Daniel.Mendez@tum.de}        
%
	 }

\date{Received: date / Accepted: date}

\maketitle

\begin{abstract}
Requirements Engineering (RE) has received much attention in research and practice due to its importance to software project success. Its interdisciplinary nature, the dependency to the customer, and its inherent uncertainty still render the discipline difficult to investigate. This results in a lack of empirical data. These are necessary, however, to demonstrate which practically relevant RE problems exist and to what extent they matter.  Motivated by this situation, we initiated the \emph{Naming the Pain in Requirements Engineering} (NaPiRE) initiative which constitutes a globally distributed, bi-yearly replicated family of surveys on the status quo and problems in practical RE. 

In this article, we report on the qualitative analysis of data obtained from 228 companies working in 10 countries in various domains and we reveal which contemporary problems practitioners encounter. To this end, we analyse 21 problems derived from the literature with respect to their relevance and criticality in dependency to their context, and we complement this picture with a cause-effect analysis showing the causes and effects surrounding the most critical problems.

Our results give us a better understanding of which problems exist and how they manifest themselves in practical environments. Thus, we provide a first step to ground contributions to RE on empirical observations which, until now, were dominated by conventional wisdom only.

\end{abstract}

\section{Introduction}
\label{sec:Intro}
Requirements Engineering (RE) aims at the elicitation, analysis, and specification of requirements that unambiguously reflect the intended purpose of a software system considering and aligning the viewpoints of all relevant stakeholders. Precise and consistent requirements directly contribute to appropriateness and cost-effectiveness in the development of a system~\cite{NE00} whereby RE is a determinant of productivity and (product) quality~\cite{DC06}. Yet, RE remains a discipline that is inherently complex due to the various influences in industrial environments. The process itself is driven by uncertainty as many aspects are usually not clear when setting up a project~\cite{MWLBC10}. The project setting, however, influences the choice of methods, approaches, and tools in RE as in no other software engineering discipline. This makes it impossible to standardise the discipline and to propose holistic solutions to RE. The interdisciplinary nature of the discipline and the dependency to various socio-economic and process-related factors that pervade RE make it difficult to investigate and improve~\cite{MW13}. 

Over the last years, we have observed a strong research community arise and propose a plethora of promising contributions to RE. Yet, we still know very little about the practical impact of those contributions or whether they are in tune with the practical problems they intend to address~\cite{MMFV14}. The state of empirical evidence in RE is particularly weak and dominated by, if at all, isolated case studies and small-scale studies investigating aspects that hardly can be generalised. It remains often unknown for which situations the observed effects of applying a specific method holds or what the long-term views are on cost and benefit when adopting and applying those methods. In most cases, accurate evaluations starve in the future work section of publications~\cite{CA07}. 

Theoretical and practical contributions to RE are heavily steered by conventional wisdom rather than empirical observations. In our current understanding, there are two reasons. First, we still do not exploit the full potential of empirical software engineering principles in RE~\cite{CDW12} to reveal theories and practically relevant improvement goals, and, in consequence, for evaluating contributions on the basis of clear and practically relevant hypotheses. Both, however, are a prerequisite for problem-driven research. Second -- and more important -- it is per se difficult to provide proper empirical figures that could demonstrate, for instance, problems in RE or even particular success factors~\cite{CA07}. This leads to the current situation where we still lack empirically grounded and comprehensive theories for RE. To reach this aim, we need, as a first step, to expand our knowledge about which problems exist in RE and what their causes are while covering the particularities of the project contexts for which those phenomena hold. This knowledge about the state of practice and potentially resulting problems in RE would allow us to better steer future research in a problem-driven manner.

This overall situation motivated the \emph{Naming the Pain in Requirements Engineering} (NaPiRE) initiative under the umbrella of the \emph{International Software Engineering Research Network} (ISERN). NaPiRE constitutes a globally distributed family of practitioner surveys on the status quo, problems, and their causes and effects in RE. The overall objective of NaPiRE is to establish an open knowledge base about the status quo as well as practical problems and needs in RE. In the long run, the obtained data shall support us in defining a holistic theory of RE covering a broad set of context factors for which particular phenomena hold. NaPiRE is currently run by 26 researchers from 14 countries around the world.

\subsection{Research Objective}
Our objective is to use the NaPiRE data from the 2014/15 survey run to explore which problems practitioners experience and what their causes and effects are. This shall allow us to provide a basis for steering RE research in a problem-driven manner.

\subsection{Contribution}

In this article, we contribute with an analysis of: 
\begin{compactenum}
\item RE problems practitioners experience in their project setting and an analysis of the problems with respect to their criticality for project failures, including a differentiated view according to chosen context factors such as the development process model used (agile or plan-driven). 
\item Most reported causes of the RE problems, as reported by our survey respondents, and their influence on the most cited RE problems.
\item Causes and effects, providing an overview on the survey results on causes and effects of the most critical RE problems.
\end{compactenum}

The analysis is based on data obtained from 228 companies in 10 countries. We expect our contributions to
\begin{compactitem}
\item increase the awareness of practitioners of problems reoccurring in RE and causes that lead to those problems, thereby allowing them to directly assess their own situation with respect to the state of practice, and 
\item provide an empirical foundation for researchers to base their investigations on a set of practical problems and causes.
\end{compactitem}

\subsection{Outline}
The article is organised as follows. In Section~\ref{sec:RelWork}, we describe related work and also provide information on the background of the NaPiRE initiative including previously published material. Section~\ref{sec:StudyDesign} introduces the study design including research questions and the data collection and analysis procedures. In Section~\ref{sec:Results}, we report on our study results, before concluding our article in Section~\ref{sec:Conclusions}.

\section{Related Work and Background}
\label{sec:RelWork}
In the following, we will discuss work related to our study, before introducing the NaPiRE initiative and previously published material in that context in detail. 
\subsection{Related Work}

There is a large body of research on requirements engineering in general and on specific RE methods in particular. Our study touches on the results of many of them, but we cannot discuss them all here. There exist surprisingly few comprehensive systematic literature reviews. For instance, there is a systematic review on effectiveness of requirements elicitation techniques \cite{davis2006effectiveness} and mapping studies on creativity \cite{lemos2012systematic}, requirements specification improvement methods~\cite{pekar2014improvement} as well as the empirical work on requirements specifications techniques \cite{condori2009systematic}. In the latter, the authors emphasise that most studies are experiments, and that the practitioners' view is missing.

We will focus on related work which performed survey research in the area of requirements engineering or at least with a strong RE component.\footnote{Parts of the following text are based on our related work discussion in \cite{MW14} as the related work has not changed significantly.} In RE survey research, we see two major areas: investigations of techniques and methods and investigations of general practices and contemporary issues in practice. Both areas investigate to some degree problems in RE and their causes.

Contributions that investigate techniques and methods analyse, for example, selected requirements phases and 
which techniques are suitable to support typical tasks in those phases. 
Cox et al.~\cite{CNV09} performed a broader investigation of all phases to analyse the perceived value of the RE practices recommended by Sommerville and Sawyer~\cite{Sommerville1997}. 
Studies like those reveal the effects of given techniques when applying them in practical contexts. 

Surveys on general practices and phenomena in industry include the well-known Chaos Report of the Standish Group, 
examining especially root causes for project failures of which most are to be seen in RE, such as missing user involvement. 
Whereas the report is known to have serious flaws in its design negatively affecting the validity of the results~\citep{EV10}, 
other studies, such as the (German) Success~study~\cite{success07}, conduct a similar investigation of German companies 
including a detailed and reproducible study design. Still, both surveys exclusively investigate failed projects and general 
causes at the level of overall software processes. 
A similar focus, but exclusively narrowed down to the area of RE, had the study of Kamata et al.~\cite{IT07}. They analysed the 
criticality of the single parts of the IEEE software requirements specification Std.~830-1998~\cite{IEEE830} on project success. 

The focus of those studies, however, does not support the investigation of contemporary phenomena and problems of RE in industry. 
Nikula et al.~\cite{nikula2000sps} present a survey on RE at the organisational level of small and medium-sized companies in Finland. Based on their findings, they inferred improvement goals, e.g., on optimising knowledge transfer. 
Staples et al.~\cite{SNJABR07} conducted a study investigating the industrial reluctance on software process improvement. 
They discovered different reasons why organisations do not adopt normative improvement solutions, for example, CMMI 
and related frameworks (focussing on assessing and benchmarking companies rather than on problem-driven improvements, 
see~\cite{NMJ09,PIGO08}). Exemplary reasons for a reluctance to normative improvement frameworks 
were the small company size where the respondents did not see clear benefit. 

A survey that directly focused on discovering problems in practical settings was performed by Hall et al.~\cite{HBR02}. 
They empirically underpin the problems discussed by Hsia et al.~\cite{HDK93} and investigated a set of critical organisational 
and project-specific problems, such as communication problems, inappropriate skills or vague requirements.
Solemon, Sahibuddin, and Abd Ghani~\cite{solemon2009requirements} report on a survey on RE problems and practices in 
Malaysian software companies. They found several of the RE problems we also saw in our survey. 
Liu, Li, and Peng~\cite{liu2010requirements} describe a survey conducted in China about practices and problems in RE. They
discuss several problems we also investigated but concentrate on China.
Verner et al.~\cite{verner2007requirements} ran a survey in Australia and the USA. They concentrated on success factors
in RE and found good requirements, customer/user involvement, and effective requirements management to be
the best predictors of project success. In the study reported in this article, we identified problems and their causes which can be used to refine the abstract success factors identified by Verner et al. For instance, we identified incomplete and/or hidden requirements as a main problem to reach 'good requirements' and weak qualification as well as lack of experience as its main causes. These causes are useful guidelines to reach more effective requirements management. Further studies on RE problems and causes are still rare. For instance, Al-Rawas and Easterbrook \cite{al1996communication} present a field study on communication problems in requirements engineering. 

In summary, we have existing work using survey research to understand specific RE techniques as well as to
understand practical problems. Yet, so far there has not been a large, replicated and world-wide analysis of RE problems
together with their causes in practice.

\subsection{The NaPiRE Initiative}
\label{sec:NaPiREInitiative}
The NaPiRE (Naming the Pain in Requirements Engineering) initiative was started by Daniel M\'endez Fern\'andez
and Stefan Wagner in 2012 as a reaction to the lack of a general empirical basis for requirements engineering research.
The basic idea was to establish a broad survey investigating the status quo of requirements engineering in practice
together with contemporary problems practitioners encounter. This should lead to the identification of interesting
further research areas as well as success factors for RE.

The initial team was convinced that because of the diversity of RE in research and practice, we would not be able to achieve
this high goal by ourselves and in a single survey. Therefore, NaPiRE was created as a means to collaborate with
researchers from all over the world to conduct the survey in different countries. This allows us to investigate RE
in different cultural environments and increase the overall sample size. Furthermore, we decided to run the survey
every two years so that we can cover slightly different areas over time and have the possibility to observe trends.
The conduct of NaPiRE is guided by the four principles described in Tab.~\ref{tab:principles}.

\begin{table}[htb]
\caption{Guiding principles of NaPiRE\label{tab:principles}}
\begin{tabular}{r p{0.65\linewidth}}
\toprule
\textbf{Openness} & Openness begins by cordially inviting researchers and practitioners of any software-engineering-related community 
to contribute to NaPiRE and ends by disclosing our results and reports without any restrictions or commercial interest. \\ 
\textbf{Transparency} & All results obtained from the distributed surveys are committed to the PROMISE repository. This shall allow 
other researchers an independent data analysis and interpretation.\\ 
\textbf{Anonymity} & The participation in NaPiRE in the form of a survey respondent is possible by invitation only. This 
supports a transparent result set and response rate. We collect no personal data, however, and every 
data set obtained from the survey will be carefully cleansed of information that might be traced back to a specific 
company to ensure that no personal data will be disclosed to the public. That is, we guarantee that no answer set 
can be related to survey participants.\\ 
\textbf{Accuracy and Validity} & With accuracy and validity, we refer in particular to the data collection and to the data 
analysis. Each question in the survey is carefully defined according to a jointly defined theory to specifically confirm 
or refute existing expectations. The data analysis is furthermore performed in joint collaboration by different researchers 
to maximise the validity of the results.\\ 
\bottomrule
\end{tabular}
\end{table}

At present, the NaPiRE initiative has members from 14 countries mostly from Europe but also North-America, South-America
and Asia. There have been two runs of the survey so far. The first was the test run performed only in Germany and in the Netherlands in 2012/13. 
The second run was performed in 10 countries in 2014/15. All up-to-date information on NaPiRE together with links
to all publications and the data is available on the web site \url{http://www.re-survey.org}.

The first run in Germany together with the overall study design was published in \cite{MW14}. A preliminary version
was also published in \cite{MW13} and the detailed data and descriptive analysis is available as technical report~\cite{MW13b}.
This first run already covered the spectrum of status quo and problems. It had additional questions on the expectations
on a good RE which we removed in the second run because they provided the least interesting insights. The
study design was described with the bi-yearly replications and world-wide distribution in detail. Furthermore, a first version
of a theory of the status quo and problems in RE was provided in the form of hypotheses. Overall, we were able to get full responses from
58 companies to test the theory. Most of the proposed theory could be supported and changes were discussed based on the data.
Finally, a detailed qualitative analysis of the experienced problems and how they manifest themselves was made. The
article at hands concentrates on the replication of that part of the survey, with further emphasis on the problems and their causes.

For the second run, we have published three papers so far, concentrating on specific aspects and the data from only one
or two countries. So far, there is no comprehensive analysis of problems and causes based on the complete, international
data set.

In \cite{kalinowski:seke15}, we used the Brazilian data to explore how to analyse problems and causes in RE in detail. In particular, we tested the use of \emph{probabilistic cause-effect diagrams} on this data to better understand the relationship of causes and problems. We introduced those diagrams for causal analysis purposes~\cite{KTC08}, and have further detailed them subsequently~\cite{KMT11}. We decided to employ these diagrams in our subsequent efforts, including this article, in which we use them for analysing causes and effects of problems based on the complete data set.

Thereafter, in \cite{mendez:softw15}, we concentrated on analysing the similarities and differences in the experienced problems between Brazil and Germany. We also used the probabilistic cause-effect diagrams for problem and cause analysis. Our key insights in this article were that the dominating factors are related to human interactions independent of country, project type or company. Furthermore, we observed a higher inclination to standardised development process models in Brazil and slightly more non-agile, plan-driven RE in Germany.

Finally, in \cite{kalinowski:swqd16}, we concentrated on the often mentioned problem of \emph{Incomplete and / or hidden requirements}
and investigated common causes for this problem based on the Austrian and Brazilian data. The most common causes we found were \emph{Missing qualification of RE team members}, \emph{Lack of experience}, \emph{Missing domain knowledge}, \emph{Unclear business needs} and \emph{Poorly defined requirements}.

\section{Study Design}
\label{sec:StudyDesign}
The overall objective of the NaPiRE endeavour is to use survey research in a globally distributed an replicated manner to build a holistic theory of the industrial status quo in requirements engineering. To this end, we conduct the survey bi-yearly while continuously adapting our questionnaire in response to data obtained from previous years (see also Sect.~\ref{sec:NaPiREInitiative}). 

In the following, we introduce those information on the study design relevant to the analysis presented of this article. Details on the overall principles and process followed in NaPiRE, as well as on the team involved, can be taken from our project website \url{RE-Survey.org}. There, we also publish the full instrument used under the publication section. 

\subsection{Research Questions}
Our objective is to get a better understanding of which problems practitioners encounter in RE, and how those problems relate to the overall project setting (causes and problems). To this end, we formulate three research questions, shown in Table~\ref{tab:rqs}, to steer the design of our study.

\begin{table}[htb]
\caption{Research questions.\label{tab:rqs}}
\begin{tabular}{p{0.13\linewidth}p{0.81\linewidth}}
\toprule
\textbf{RQ~1} & Which contemporary problems exist in RE? \\ 
\textbf{RQ~2} & What are observable patterns of problems and context characteristics?\\
\textbf{RQ~3} & What are their perceived causes and effects? \\
\bottomrule
\end{tabular}
\end{table}

The first question aims at understanding which problems practitioners experience in general in their RE and what their criticality is w.r.t.\ project failure. This more descriptive view is complemented by the second research question, which aims at understanding whether there exist problems that relate to specific context factors, such as the company size or the type of used process model. Once we understand whether there exist specific patterns in the problems, we want to know what their perceived causes and implications are going beyond project failure.

\subsection{Instrument}

The overall instrument used in NaPiRE constitutes in total 35 questions used to collect data on (a) the demographics, (b) how practitioners elicit and document requirements, (c) how requirements are changed and aligned with tests, (d) what and how RE standards are applied and tailored, (e) how RE is improved, and finally (f) what problems practitioners experience in their RE. In the study at hands, we focus on the problems practitioners experience in their RE while using the answers given to selected questions on the status quo to answer RQ~2. Table~\ref{tab:instrument} summarises an excerpt of our questionnaire demonstrating the scope of our study.  The full questionnaire can be taken from the publication section of our project website.

\begin{table*}[tb]
\centering \scriptsize
\caption{Questions (simplified and condensed excerpt).}
\label{tab:instrument}
\begin{tabular}{crp{0.55\linewidth}l}
\toprule
\textbf{Parts}& \textbf{No.}& \textbf{Question} & \textbf{Type} \\ \hline 
Demographics           & Q~1        & What is the size of your company? & Closed(SC)\\
            & Q~2           & Please describe the main business area and application domain. & Open\\
             & Q~3          & Does your company participate in globally distributed projects? & Closed(SC)\\
             & Q~4          &  In which country are you personally located? & Open \\
            & Q~5           & To which project role are you most frequently assigned? & Closed(SC) \\
            & Q~6           &  How do you rate your experience in this role? & Closed(SC)\\
            & Q~7           &  Which organisational role does your company take most frequently in your projects? & Closed(MC)\\
            & Q~8          &  Which process model do you follow (or a variation of it)? & Closed(MC)\\ \hline
Status Quo & Q~9      &  How do you elicit requirements?	 & Closed(MC)  \\
            & Q~10           &  How do you document functional requirements?	 & Closed(SC)  \\
            & Q~11           &  How do you document non-functional requirements?       & Closed(SC)  \\
            & Q~12           &  How do you deal with changing requirements after the initial release?	& Closed(SC)  \\
             & ...          &  ... & ... \\
            & Q~16           &  What requirements engineering company standard have you established at your company? & Closed(MC) \\
                & ...          &  ... & ... \\ \hline
Problems  & Q~28            &  Considering your personal experiences, how do the following (more general) problems in requirements engineering apply to your projects? & Likert  \\
              & Q~29         &  Considering your personally experienced problems (stated in the previous question), which ones would you classify as the five most critical ones (ordered by their relevance). & Closed \\
            & Q~30           &  Considering your personally experienced most critical problems (selected in the previous question), which causes do they have? & Open \\
           & Q~31       &  Considering your personally experienced most critical problems (selected in the previous question), which implications do they have? & Open \\

                & Q~32       &  Considering your personally experienced most critical problems (selected in the previous question), which mitigations do you define (if at all)? & Open \\
                & Q~33       &  Considering your personally experienced most critical problems (selected in the previous question), which would you classify as a major cause for project failure (if at all)? & Closed(MC) \\ \bottomrule
\end{tabular}
\end{table*}

In this part, we use a mix of open questions and closed ones. The type of question is denoted in the table (last column). In case of closed questions, the answers can be mutually exclusive single choice answers (SC) or multiple choice ones (MC). Most of the closed multiple choice questions include a free text option, e.g., ``other'' so that the respondents can express company-specific deviations from standards we ask for. We furthermore use Likert scales on an ordinal scale of 5 and define for each a maximum value (e.g., ``agree'', or ``very important''), a minimum value (e.g., ``disagree'', or ``very unimportant''), and the middle (``neutral''). The latter allows the respondents to make a selection when they have, for example, no opinion on the given answer options. Finally, we define selected questions as conditional ones to guide through the survey by filtering subsequent question selection. For instance, if respondents state that they have not defined any company-specific RE standard (Q~16), the remaining questions on the standards are omitted. 

For the analysis of the problems (Q~28 to Q~33), we first present a list of problems practitioners are meant to typically encounter in practice. This list emerged from previously conducted external studies (see also our related work section~\ref{sec:RelWork}) and has been already used in our first survey round (see also~\cite{MW14} for a richer discussion on the elaboration of the list). For this survey round, we use the same list which includes a set of 21 pre-compiled problems shown next in no particular order:
\begin{compactitem}
\item Communication flaws within the project team
\item Communication flaws between project team and the customer
\item Terminological problems
\item Unclear responsibilities
\item Incomplete and / or hidden requirements
\item Insufficient support by project lead
\item Insufficient support by customer
\item Stakeholders with difficulties in separating requirements from previously known solution designs
\item Inconsistent requirements
\item Missing traceability
\item Moving targets (changing goals, business processes and / or requirements)
\item Gold plating (implementation of features without corresponding requirements)
\item Weak access to customer needs and / or (internal) business information
\item Weak knowledge of customer's application domain
\item Weak relationship to customer
\item Time boxing / Not enough time in general
\item Discrepancy between high degree of innovation and need for formal acceptance of (potentially wrong / incomplete / unknown) requirements
\item Technically unfeasible requirements
\item Underspecified requirements that are too abstract and allow for various interpretations
\item Unclear / unmeasurable non-functional requirements	
\item Volatile customer's business domain regarding, e.g., changing points of contact, business processes or requirements
\end{compactitem} 

The respondents are then asked to report the relevance of the presented problems for their project setting, before being asked to select the 5 most critical ones (Q~29). The subsequent questions on the causes, the effects, and potential mitigation strategies consider then those 5 problems.  

\subsection{Data Collection}
The survey is conducted by invitation only to have a better control over the distribution of the survey among specific companies and also to control the response rate. The responses where, however, anonymous to allow our respondents to freely share their experiences made within their respective company. For each company, we invited one respondent as a representative of the company. In case of large companies involving several autonomous business units working each in a different industrial sector and application domain, we selected a representative of each unit. For the data collection, each country representative defined an invitation list including contacts from different companies and initiated the data collection independently as an own survey project. All surveys relied on the same survey tool\footnote{We implemented the survey as a Web application using the \emph{Enterprise Feedback Suite}.} hosted and administrated by the representatives for Germany. Information on the overall data collection procedure can be also taken from our project website.

For the study at hands, we conducted the survey in the countries summarised in Table~\ref{tab:datacollection}. The data collection phase varied among the countries and some of the countries also collected the data in multiple tranches potentially resulting in longer inactivity phases.

\begin{table}[htb]
\small
\centering
\caption{Data collection phase (overview).}
\label{tab:datacollection}
\begin{tabular}{lll}
\toprule
\textbf{Area}  & \textbf{Country} & \textbf{Data Collection Phase} \\ \hline
Central Europe (CE)	&	&			\\
&	Austria (AT)	&	 2014-05-07 to 2014-09-15		\\
&	Germany (DE)	&	2014-05-07 to 2014-08-18		\\
& Ireland (IE) & 2014-05-07 to 2014-12-31 \\
North America (NA)	&	&			\\
& Canada	(CA)	&	2014-05-07 to 2015-08-15		\\
& United States of America (US) 	&	2014-05-07 to 2015-05-01		\\
Northern Europe (NE)	&	&			\\
& Estonia	(EE)	&	2014-05-07 to 2014-10-31		\\
& Finland	(FI)	&	2015-06-01 to 2015-08-28		\\
& Norway	(NO)	&	2014-05-07 to 2014-09-15		\\
& Sweden	(SE)	&	2014-05-07 to 2014-09-15		\\
South America (SA)	&	&			\\
& Brasil (BR)	&	2014-12-09 to 2015-03-31		\\ \bottomrule
\end{tabular}
\end{table}

\subsection{Data Analysis}
\label{sec:DataAnalayis}
To answer our research questions, we first need to quantify the answers given for the selection of the predefined problems the participants shall rank as they have experienced them in their projects. As part of this quantification, we also sum up to which extent the given problems have led to project failures in the experience of the participants. 

For analysing the answers given to the open question on what causes and effects the RE problems have (Q~30 to Q~32), we apply qualitative data analysis techniques as recommended in context of Grounded Theory~\cite{GS67, AHK11}. In particular, we considered the following basic coding steps: 
\begin{compactenum}
\item \emph{Open coding} to analyse the data by adding codes (representing key characteristics) to small coherent units in the answers, and categorising the developed concepts in a hierarchy of categories as an abstraction of a set of codes -- all repeatedly performed until reaching a state of saturation. We define the (theoretical) saturation as the point where no new codes (or categories) are identified and the results are convincing to all participating researchers~\cite{BM11}.
\item \emph{Axial coding} to define relationships between the concepts, e.g., ``causal conditions'' or ``consequences''.
\item \emph{Internal Validation} as a form of internal quality assurance of the obtained results.
\end{compactenum}

Please note that we deviate from the approach to Grounded Theory as introduced by Glaser and Strauss~\cite{GS67} in two ways. First, given that we analyse data from an anonymously conducted survey after the fact, we are not able to follow a constant comparison approach where we iterate between the data collection and the analysis.  This also means that we are not able to validate our results with the participants, but have to rely in internal validation procedures to reduce the resulting threats to the validity (see also Sect.~\ref{sec:ValidityProcedures} discussing our validity procedures). Second, we do not inductively build a theory from bottom up, as we start with a predefined conceptual model (i.e. the problems) whereby we do not apply selective coding to infer a central category.

In our instrument, we have already a predefined set of RE problem codes (given RE problems, see Q~28 and 29) for which we want to know how the participants see their causes and effects. For this reason, we rely on a mix of bottom-up and top-down approach. That is, we start with our pre-defined core category, namely \emph{RE problems} and a set of codes each representing one key RE problem, and three sub-categories: \emph{Causes}, \emph{Effects}, and \emph{Mitigation Strategies}, which then group the codes emerging from the free text answers given by the participants. Within the causes and effects, we pre-define again the sub-categories. These sub-categories are as follows:
\begin{compactitem}
\item For the causes, we use the sub-categories \emph{Input}, \emph{Method}, \emph{Organization}, \emph{People}, \emph{Tools} suggested in our previous work on defect causal analysis ~\cite{KCT12} as we want to know from where in the socio-economic context the problems stem.
\item For the implications, we use the sub-categories \emph{Customer}, \emph{Design or Implementation}, \emph{Product}, \emph{Project or Organization}, and \emph{Verification or Validation} as done in our previous work~\cite{mendez:softw15} as we want to know where in the software project ecosystem the problems manifest themselves.
\end{compactitem}

For each answer given by the participant, we then apply open coding and axial coding until reaching a saturation in the codes and relationships, and we allocate the codes to the previously defined sub-categories. A rich discussion on the principles of analysing textual data and how we generally apply it in context of the NaPiRE initiative can also be found in our previously published material~~\cite{WM15}. 

For coding our results, we first coded in a team of two coders the first 250 statements to get a first impression of the resulting codes, the way of formulating them and the level of abstraction for capturing the codes. After having this overview, we organised a team of five coders within Germany and Brazil. Each of the coders then coded approx.~200 statements for causes and additional ~200 statements for effects, while getting the initial codes from the pilot phase as orientation. In case the coder was not sure how to code given statements, she marked the code accordingly for the validation phase. During that validation phase, we formed an additional team of three independent coders who then reviewed those codes marked as ``needs validation'' as well as an additional sample, comprising 20\% of the statements assigned to each coder, selected on their own. After the validation phase, we initiated a call where we discussed last open issues regarding codes which still needed further validation, before closing the coding phase. The overall coding process took in total three months.

To interpret the resulting codes, in particular the answers to research question 2 where we want to know whether there exist patterns of problems and context characteristics, we need to put the answers to the problems in relation to the answers given to previous questions in the questionnaire. To this end, we apply manual blocking to our results, i.e. we select subsets of results which include specific variable selections; for instance all results for which a specific process model has been selected. We then discuss in the group of researchers whether there are specific differences in the problems visible, e.g. compared to the problems when other process models have been selected. However, blocking the codes according to all possible permutations of the variables in the questionnaire is not feasible. For this reason, we intentionally block the codes according to a combination of the two variables \emph{company size} (Q~1) and the type of \emph{software process models} used (Q~8) (agile or plan-driven), which we believe to be suitable for an initial observation of relevant patterns. Of course, further blocking variables from the status quo section of the questionnaire could be used. However, the relation of the whole underlying NaPiRE theory to the manifestation of the problems also involves significant effort and is left for future work.

\subsection{Validity}
\label{sec:ValidityProcedures}
The overall NaPiRE endeavour includes several procedures for checking validity, i.e., concerning the data collection and analysis phases, as described in detail in our previously published material~\cite{MW14}. For the analysis of qualitative data, which is in the scope of this article, we defined additional procedures as described next. 

The major threat to validity arises from the actual coding process as coding is essentially a creative task. Subjective facets of the coders, such as experiences, expertise, and expectations, strongly influence the way we code free text statements. A further threat arises from the fact that we cannot validate our results with the respondents given the anonymous nature of our survey. Finally, coding over a distributed team of researchers can additionally lead to a possibly limited degree of saturation in the emerging codes. 

To decrease the threats, we first conducted a pilot phase in the analysis. After agreeing on the first resulting codes within the group of coders, i.e. after getting a common understanding on the wording and the level of abstraction in the codes, we then applied researcher triangulation for the actual coding process. An independent group of researchers coded all the statements given by the respondents, before we subsequently conducted a validation phase within the group of researchers. The validation phase during coding should then minimise the amount of incorrect codes. This validation focussed on codes declared as ``needs validation'', but also on further codes presumably being coded correctly. There, we focused on the occurrences of the codes rather than on the choice of the labels for the codes (e.g., “CRs” and “change requests” are seen as the same code). 

\section{Study Results}
\label{sec:Results}
In the following, we present the survey results concerning the RE problems (RQ~1), observable patterns (RQ~2) and their common causes and effects (RQ~3) as reported by our respondents. We first summarise the information about the study population, characterising the responding organisations, as this information is crucial to enable a suitable interpretation of the results.

\subsection{Study Population}
\label{sec:DataPopulation}
In total, 354 organisations spread across 10 different countries agreed to answer the survey. Out of these, 228 (63\%) completed the survey by going through all of its questions and successfully reaching its end. Table~\ref{tab:pupulation} shows the number of completed datasets and the completion rate per country.

\begin{figure}[!hbtp]
\centering
  \includegraphics[width=0.8\textwidth]{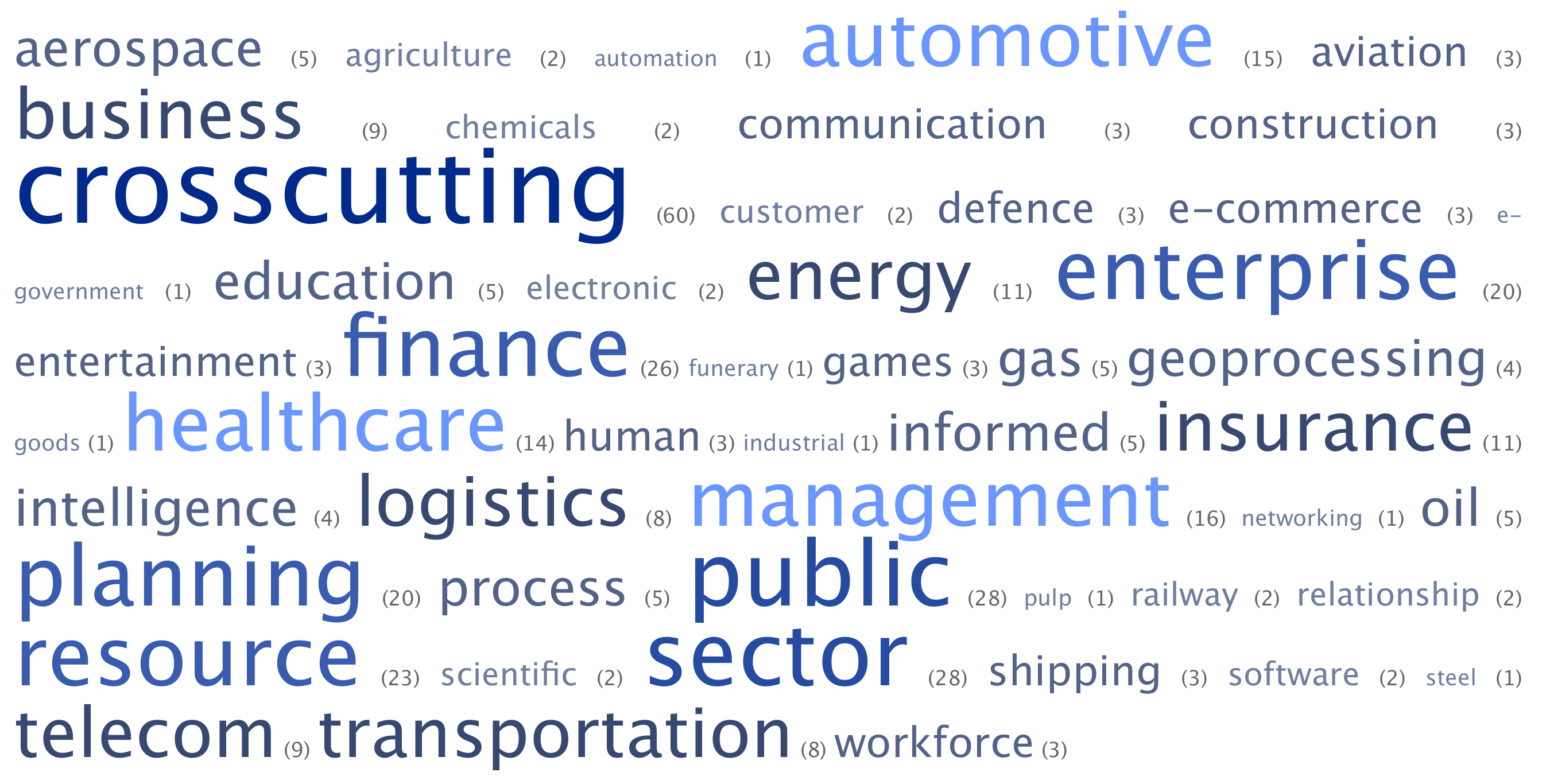}\\
  \caption{Tag cloud of the business domains of the responding organisations.}
  \label{fig:BusinessDomains}
\end{figure}

\begin{table}[htb]
\small
\centering
\caption{Study population including response rates, total datasets obtained, completed datasets, and completion rates. The explanation of the country codes can be take from Table~\ref{tab:datacollection}.}
\label{tab:pupulation}
\begin{tabular}{cccccc}
\toprule
 &  & \textbf{Response} & \textbf{Total} & \textbf{Completed} & \textbf{Completion}  \\
\textbf{Area}  & \textbf{Country} & \textbf{Rate} & \textbf{Datasets} & \textbf{Datasets} & \textbf{Rate}  \\ \hline
CE	&	&		&		&	&	\\
&	AT	&	72\% & 18	&	14	&	78\% 	\\
&	DE	&	36.8\% & 50	&	41	&	82\%  \\
&	IE	&	39.7\% & 25	&	18	&	72\%  \\
NA	&	&	&	&		&	 	\\
&	CA	&	75\% & 15	&	13	&	87\% \\
&	US	&	36.2\% & 25	&	15	&	60\% \\
NE	&	&		&		&	 &	\\
&	EE	&	25.7\% &9	&	8	&	89\%	\\
&	FI	&	37.5\% &	18 &	15	&	83\% \\
&	NO	&	70.8\% &	17 &	10	&	59\% \\
&	SE	&	 51.8\% & 59	&	20	&	34\% \\
SA	&	&		&		&	& \\
&	BR	&	35.3\% & 118	&	74	&	63\% \\ \hline
&  & Total: & 354 & 228 & 64\% \\
\bottomrule
\end{tabular}
\end{table}
The results reported in this article consider the completed datasets only. These 228 organisations were active in a variety of different business domains. The domains were provided by the respondents in free text format (see  Table~\ref{tab:instrument}, question Q2) and coded by the researchers. The tag cloud for the coded business domains can be seen in Figure~\ref{fig:BusinessDomains}. 

This figure shows the frequency of each domain code and highlights the most frequent ones. At all, 215 of the 228 organisations provided answers for their business domain. There is a huge variety in the business domains ranging from embedded software systems (described by the respondents as ``Automotive, Embedded Software'' or ``Software for medical devices'') over business systems (``business intelligence for data centres'' or ``Software ERP'') to consulting (``IT Consulting'' or  ``Consulting for secure systems and software''). The most frequent code assigned was \emph{cross-cutting} which means that the organisation is actively  working with products and/or services that can be applied to several different business domains (e.g., cloud computing and web applications, custom software development, enterprise resource planning products, IT consulting services). Additionally, we identified a very large amount of different domains with few data points in each one. Therefore, considering the amount of organisations active in several domains and the large variety of different domains reported, we decided to characterise the responding organisations independently of their business domain, in terms of domain cross-cutting characteristics of size and process model used (see also Sect.~\ref{sec:DataAnalayis}).

Concerning size, we grouped organisations as small, medium, and large-sized. For this grouping, as in~\cite{mendez:softw15}, we used the number of employees (software and other areas). Organisations with up to 50 employees were considered small-sized organisations, with 51 to 250 were considered medium-sized organisations, and with more than 250 were considered large-sized. Out of the 228 organisations that completed the questionnaire, 216 provided their number of employees. The sizes of these organisations are shown in Table~\ref{tab:organisationsizes}. 

\begin{table}[htb]
\scriptsize
\centering
\caption{Sizes of responding organisations.}
\label{tab:organisationsizes}
\begin{tabular}{lcccccc}
\toprule
& \textbf{Central} & \textbf{North} & \textbf{Northern} & \textbf{South} & \\
\textbf{Size}  & \textbf{Europe} & \textbf{America} & \textbf{Europe} & \textbf{America} & \textbf{Total} \\ \hline
Small & 12 & 11 & 20 & 26 & 69 \\
Medium & 4 & 0 & 12 & 17 & 33 \\
Large & 36 & 16 & 34 & 28 & 114 \\ \hline
Total & 52 & 27 & 66 & 71 & 216 \\
\bottomrule
\end{tabular}
\end{table}

We can observe that the datasets include relatively large samples of both, small and large-sized organisations. Considering the distributions of size per region, except for SA, the responding large-sized organisations slightly outweigh the small and medium-sized organisations.

Regarding the process models used, respondents answered a multiple choice question with the following options: RUP, Scrum, V-Model XT, Waterfall, XP, and Other (in this case informing textually which process model they use). We grouped these process models into \emph{agile} (Scrum and XP), \emph{plan-driven} (RUP, Waterfall and V-Model XT), and \emph{mixed} (for those organisations that informed to use agile and plan-driven process models or variations therein). Out of the 228 organisations that completed the questionnaire, 196 selected one of the five predefined options for their process model. The process model of these organisations is shown in Table~\ref{tab:swprocessmodels}. 

\begin{table}[htb]
\scriptsize
\centering
\caption{Software process models used in responding organisations.}
\label{tab:swprocessmodels}
\begin{tabular}{lcccccc}
\toprule
& \textbf{Central} & \textbf{North} & \textbf{Northern} & \textbf{South} & \\
\textbf{Model}  & \textbf{Europe} & \textbf{America} & \textbf{Europe} & \textbf{America} & \textbf{Total} \\ \hline
Agile & 12 & 13 & 35 & 32 & 92 \\
Plan-driven & 15 & 4 & 8 & 19 & 46 \\
Mixed & 17 & 8 & 19 & 14 & 58 \\ \hline
Total & 45 & 24 & 62 & 65 & 196 \\
\bottomrule
\end{tabular}
\end{table}

Again, the datasets include relatively large samples of both, agile and plan-driven organisations. Considering the distribution per region, except for CE, the responding organisations following agile process models in the respondents environment outweigh the plan-driven ones. The amount of organisations using mixed process models (or more than one) is large. However, we decided to exclude the mixed ones from our corresponding analyses to remove a potential confounding factor, as in these cases we had no information on the extent to which each process model is applied in the organisation.

As we believe that agile and plan-driven process models may have different effects on small, medium, and large-sized organisations, we also crossed these two characterisation dimensions aiming at further exploring potential RE problem patterns (cf. Section~\ref{sec:ResultsRQ2}). The result of this crossing is shown in Table~\ref{tab:swprocessmodelsandsize}. 

\begin{table}[htb]
\scriptsize
\centering
\caption{Responding organisations by size and process models used.}
\label{tab:swprocessmodelsandsize}
\begin{tabular}{llccccc}
\toprule
\textbf{Process}& & \textbf{Central} & \textbf{North} & \textbf{Northern} & \textbf{South} & \\
\textbf{Model}& \textbf{Size}  & \textbf{Europe} & \textbf{America} & \textbf{Europe} & \textbf{America} & \textbf{Total} \\ \hline
Agile & Small & 2 & 4 & 10 & 14 & 30 \\
Agile & Medium & 2 & 0 & 10 & 10 & 22 \\
Agile & Large & 9 & 8 & 14 & 8 & 39 \\
Plan-driven & Small & 3 & 2 & 1 & 5 &11 \\
Plan-driven & Medium & 1 & 0 & 1 & 2 & 4 \\
Plan-driven & Large & 10 & 2 & 6 & 12 & 20 \\ \hline
& Total & 27 & 16 & 42 & 51 & 136 \\
\bottomrule
\end{tabular}
\end{table}

Excluding the 58 organisations with mixed process models, 136 organisations that completed the questionnaire informed the number of employees and predefined process model options. While agile process models are being applied by small and large-sized representatives of the responding organisations, plan-driven process models are mainly applied by the large-sized ones (although we still have some samples of small sized plan-driven organisations).

We therefore could obtain a balanced characterisation of small, medium and large-sized organisations of different regions enrolled in both, plan-driven and agile development methods.

\subsection{Problems in RE (RQ~1)}
\label{sec:ResultsRQ1}

\begin{figure}[!hbtp]
\centering
  \includegraphics[width=0.8\textwidth]{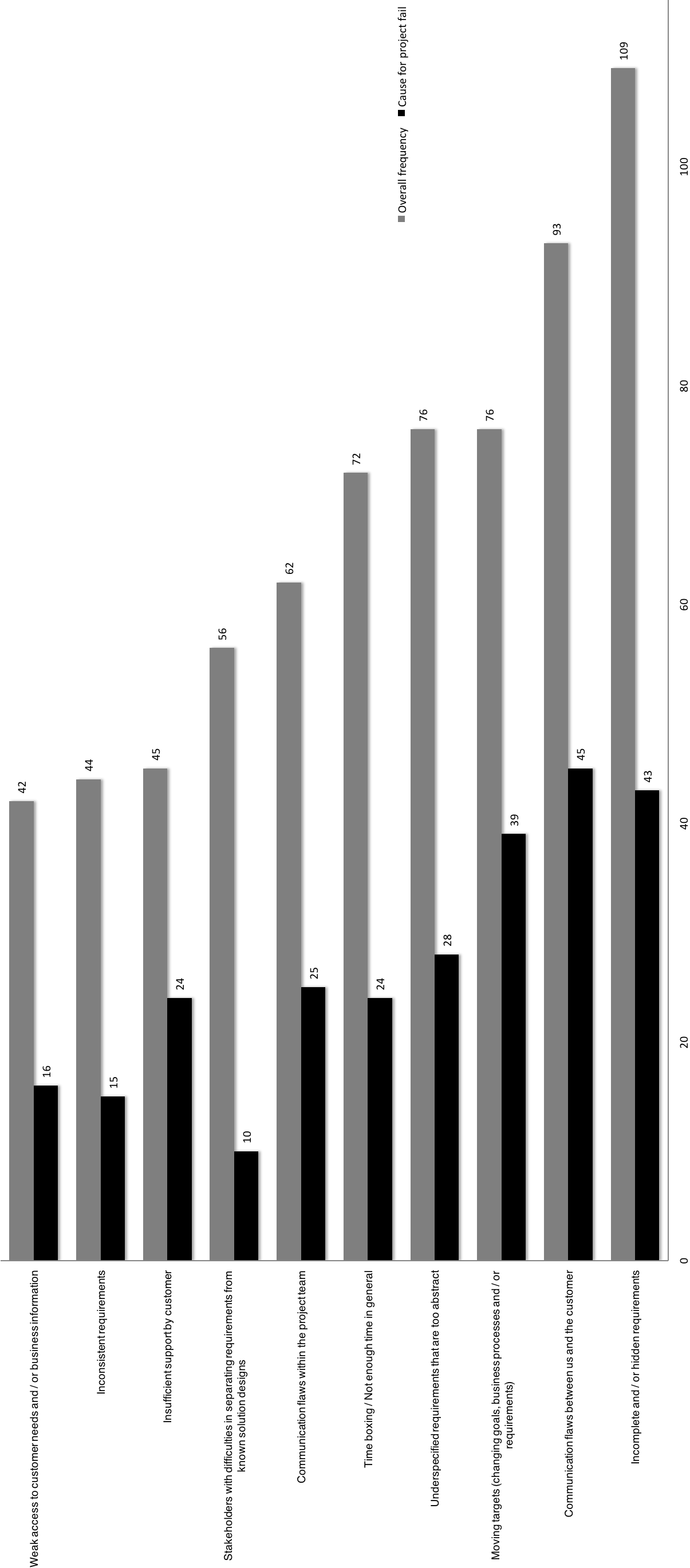}\\
  \caption{Overall frequency of top 10 RE problems and their relation to project failure.}
  \label{fig:REProblems}
\end{figure}

Based on the set of 21 pre-defined general RE problems listed in the NaPiRE questionnaire, the respondents were asked to rank the five most critical ones. The top 10 most critical RE problems, as informed by the 228 respondents, are visualised together with the frequency in which they are meant to lead to project failure in a simplified manner in Figure~\ref{fig:REProblems}. 

Table~\ref{tab:mostcriticalreproblems} further summarises the 10 most cited RE problems providing more details. There, we report on how many of the respondents cited particular problems, how many considered them as a major cause for project failure, and how often each problem was ranked in each of the five potential ranking positions, thus, showing how the bars in Figure~\ref{fig:REProblems} are composed.

\begin{table}[htb]
\scriptsize
\centering
\caption{Most cited top 10 RE problems.}
\label{tab:mostcriticalreproblems}
\begin{tabular}{p{0.45\linewidth}cccccccc}
\toprule
\textbf{RE Problem}  & \textbf{Total} & \rot{\textbf{Cause for project failure}} & \rot{\textbf{Ranked as \#1}} & \rot{\textbf{Ranked as \#2}} & \rot{\textbf{Ranked as \#3}} & \rot{\textbf{Ranked as \#4}} & \rot{\textbf{Ranked as \#5}}\\ \hline
Incomplete and / or hidden requirements	&	109	(48\%) & 43	&	34	&	25	&	23	&	17	&	10\\
Communication flaws between project team and customer	 &	93 (41\%)	&	45	&	36	&	22	&	15	&	9	&	11\\
Moving targets (changing goals, business processes and / or requirements)	&	76 (33\%)	&	39	&	23	&	16	&	13	&	12	&	12\\
Underspecified requirements that are too abstract	&	76 (33\%)	&	28	&	10	&	17	&	18	&	19	&	12\\
Time boxing / Not enough time in general	&	72 (32\%)	&	24	&	16	&	11	&	14	&	17	&	14\\ 
Communication flaws within the project team	&	62 (27\%)	&	25	&	19	&	13	&	11	&	9	&	10\\
Stakeholders with difficulties in separating requirements from known solution designs	&	56 (25\%)	&	10	&	13	&	13	&	12	&	9	&	9\\ 
Insufficient support by customer	&	45 (20\%)	&	24	&	6	&	13	&	12	&	6	&	8\\ 
Inconsistent requirements	&	44 (19\%)	&	15	&	8	&	9	&	6	&	9	&12\\
Weak access to customer needs and / or business information	&	42 (18\%)	&	16	&	7	&	10	&	8	&	8	&9\\
\bottomrule
\end{tabular}
\end{table}

Out of these critical RE problems we highlight the first five, which were cited by more than 30\% of the respondents. Noteworthy is also, however, that some problems even if they seem not to occur as often as others, seem to be still more critical as they are meant to lead more often to project failure; for instance, \emph{Incomplete and / or hidden requirements} being the most frequently cited problem is, from a relative point of view, not meant to lead as often to project failure as \emph{Communication flaws between project team and the customer} even if it occurs more often. Furthermore, \emph{Moving targets} do lead more often to project failures than \emph{Underspecified requirements that are too abstract} even if they show the same total frequency of occurrence.

The analysis of the problem patterns described next, concentrates on the top five problems.

\subsection{Problem Patterns (RQ~2)}
\label{sec:ResultsRQ2}

Given the diversity of the responding organisations, in particular concerning their sizes and process models, we block the results according to those context factors to further investigate how the problems manifest within such clusters, aiming at identifying potential RE problem patterns (see also Sect.~\ref{sec:DataAnalayis}). Table~\ref{tab:problemsswprocessmodelsandsize} shows the five most critical RE problems per process model and organisational size. We can see that besides \emph{Gold Plating}, which appears for plan-driven medium-sized organisations, all other problems are also listed in the list of the overall 10 most cited RE problems (Table~\ref{tab:mostcriticalreproblems}). Nevertheless, it is noteworthy that this specific cluster had only 4 organisations and that this fact might not represent a relevant difference. Also the textual statements of the corresponding respondents did not show much specifics of plan-driven, medium-sized companies in that respect. Only the statement ``The team believes to be very qualified to understand the business`` hints in the direction that in a plan-driven development process, the developers are not exposed to as much customer feedback as necessary and think they already know the customer's business.

\begin{table}[htb]
\scriptsize
\centering
\caption{5 most critical problems per process model used and company size.}
\label{tab:problemsswprocessmodelsandsize}
\begin{tabular}{p{0.07\linewidth}lcp{0.52\linewidth}c}
\toprule
\textbf{Process}& & &  & \textbf{Citation} \\
\textbf{Model}& \textbf{Size}  & \textbf{Total} & \textbf{Top 5 Problems} & \textbf{Count} \\ \hline

Agile & Small & 30 &1. Incomplete and / or hidden requirements & 18 (60\%) \\
         &           &      &2. Communication flaws between project team and the customer & 14 (47\%) \\
         &           &      &3. Underspecified requirements that are too abstract & 13 (43\%) \\
         &           &      &4. Communication flaws within the project team & 10 (33\%) \\
         &           &      &5. Time boxing / Not enough time in general & 13 (43\%) \\ \hline
Agile & Medium & 20 &1. Communication flaws between project team and the customer & 12 (55\%) \\
         &           &      &2. Incomplete and / or hidden requirements & 9 (41\%) \\
         &           &      &3. Communication flaws within the project team & 8 (36\%) \\
         &           &      &4. Stakeholders with difficulties in separating requirements from known solution designs & 8 (36\%) \\
         &           &      &5. Weak access to customer needs and / or business information & 7 (32\%) \\ \hline
Agile & Large &  39&1. Incomplete and / or hidden requirements & 17 (44\%) \\
         &           &      &2. Moving targets (changing goals, business processes and / or requirements) & 17 (44\%) \\
         &           &      &3. Communication flaws between project team and the customer & 15 (38\%) \\
         &           &      &4. Time boxing / Not enough time in general & 14 (36\%) \\
         &           &      &5. Underspecified requirements that are too abstract & 11 (28\%) \\ \hline
Plan-driven  & Small & 11  &1. Incomplete and / or hidden requirements & 7 (64\%) \\
         &           &      &2. Communication flaws within the project team & 6 (55\%) \\
         &           &      &3. Moving targets (changing goals, business processes and / or requirements)& 6 (55\%) \\
         &           &      &4. Time boxing / Not enough time in general & 5 (45\%) \\
         &           &      &5. Underspecified requirements that are too abstract & 5 (45\%) \\ \hline
Plan-driven  & Medium & 4  &1. Communication flaws between project team and the customer & 2 (50\%) \\
         &           &      &2. Gold plating (implementation of features without corresponding requirements) & 2 (50\%) \\
         &           &      &3. Incomplete and / or hidden requirements & 2 (50\%) \\
         &           &      &4. Moving targets (changing goals, business processes and / or requirements) & 2 (50\%) \\
         &           &      &5. Underspecified requirements that are too abstract & 2 (50\%) \\
         &           &      &5. Weak access to customer needs and / or business information & 2 (50\%) \\ \hline
Plan-driven  & Large & 30  &1. Incomplete and / or hidden requirements & 14 (47\%) \\
         &           &      &2. Communication flaws between project team and the customer & 13 (43\%) \\
         &           &      &3.Underspecified requirements that are too abstract & 10 (33\%) \\
         &           &      &4. Communication flaws within the project team & 9 (30\%) \\
         &           &      &5. Moving targets (changing goals, business processes and / or requirements) & 8 (27\%) \\
         &           &      &5. Stakeholders with difficulties in separating requirements from known solution designs & 8 (27\%) \\
\bottomrule
\end{tabular}
\end{table}

Concerning the occurrence of the problems within the clusters, the only problem being consistently in the top 3 is \emph{Incomplete and/or hidden requirements}. It is also the most cited problem overall. We will discuss its causes and effects in detail in Sec.~\ref{sec:causes-effects-top-problems}.

Very common is also \emph{Communication flaws between project team and the customer}. It appears in the three most cited problems in all clusters 
except for plan-driven and small-sized organisations. We can see one reason in the free-text answers especially for large companies. They tend to split 
the work in several teams of which some work directly with customers, while others don't. One respondent describes that their ``sales or account teams, 
product managers [\ldots] act as proxies for the end user''. Small and agile companies seem to suffer especially from customers not willing to participate 
with a considerable amount of time (``Not enough customers willing to help out and also time constraints'', ``Customer is busy and skips meetings.'' or 
``Customers have no time to explain what they actually need''). The plan-driven,
small companies who rated this problem as important, did not show a consistent pattern of reasons in their free-text answers.

Another difference concerns the \emph{Moving targets} problem. This problem is faced by all plan-driven but also large agile organisations. That
plan-driven companies cite this problem often in comparison to agile companies supports the basic premise of agile software development that it helps
to quickly adapt to changing needs. The respondents from plan-driven organisations mention as reasons the ``Lack of change management on the
customer side'', the ``Unclear business vision and understanding by stakeholders'' and overall ``badly written requirements''. The negative effects on their projects are manifold including ``project delays; extended engagement of resources beyond original plan; 
customer dissatisfaction'' and ``expensive projects, time consuming implementation, bad quality''.

But why do also large, agile companies often experience the \emph{Moving targets} problem? We do not see a clear answer from the free-text
responses. Some of the answers could be explained such that large companies in general have larger, more complex projects which also might
run for a longer time. Then the mentioned problems are more significant. For example, a change in the management of the customers was mentioned
and seems to have large effects: ``senior management confusion/churn''. But also the chance that over time other people bring in new ideas and
constraints seems to be more likely: ``There are always some stakeholders involved in later part of the project who would come up with new things''.
Even agile development processes cannot compensate this.

Another difference between the clusters concerns the \emph{Time boxing} problem, which appears mainly in agile and in small organisations. 
In both agile and plan-driven, small companies, we found three (related) reasons for this prevalence of time boxing problems: bad estimations,
unrealistic release dates and scope changes. Our respondents often mentioned that estimations were not accurate: ``A combination of bad planning 
and bad estimation of time for development'' or ``Bad estimates, unrealistic expectations''.
Especially sales and marketing is blamed for promising unrealistic dates: ``Sales shouldn't give wishful promises'' or ``Release dates are sometimes 
arbitrary and often released early to customers creating a hard deadline''. At last, frequent scope changes seem to contribute to this problem:
``Last minute changes; change of priority; Business urgency''.

Finally, we noticed that \emph{Weak access to customer needs and / or business information} only appeared in the medium-sized clusters. For all 
medium-sized organisations, including the ones with mixed process models, it was the third most cited problem, with 13 citations, while it did not 
appear in the top 5 RE problems for small- and large-sized organisations. We could not find any consistent patterns in the free-text answers
of the medium-sized companies. We can only speculate that small-sized organisations might adapt themselves to fit the availability of their customers, 
while large organisations might have more influence on their customers to achieve the required access.


\subsection{Cause-Effect Analysis (RQ~3)}
\label{sec:ResultsRQ3}

In the following, we will summarise causes and effects as reported by our respondents for the discussed RE problems. After selecting the five most critical RE problems, we asked our respondents to provide what they believe to be the main causes and effects for each of the problems. They provided the causes and effects in an open question format, with one open question for the cause and another for the effect for each the previously selected RE problems. Details on the data analysis procedure can be taken from Section~\ref{sec:DataAnalayis}. In the following, we first discuss the main causes for the RE problems, before discussing the causes and effects for the top problems in detail.

\subsubsection{Main Causes for RE Problems}

In total, 177 of the 228 organisations that completed the survey provided textual information for at least one cause and we received in total 820 textual answers for causes and effects of RE problems. The coding process yielded 92 different codes for causes of RE problems and 49 different codes for their effects. While it does not make sense to analyse effects out of the context of the RE problems that provoke them, causes are at the beginning of the causal system~\cite{Card05}. Thus, an isolated view on causes (without consideration of their specific RE problem context) may provide valuable information, for example, on how to prevent RE problems in general. We therefore first provide a descriptive view on the most cited causes of RE problems.

The ten most reported causes and how often they have been reported within each of the analysed clusters are shown in Table~\ref{tab:topcauses}. For the percentages, we considered the total amount of organisations that completely answered the survey, given that empty answers could mean that they did not consider any specific cause.

\begin{table}[htb]
\scriptsize
\centering
\caption{Most cited causes of RE problems. Clusters yielding in more than 20\% frequency of the causes are highlighted.}
\label{tab:topcauses}
\begin{tabular}{p{0.22\linewidth}p{0.09\linewidth}llllll}
\toprule
                       &                   &  \textbf{Agile} & \textbf{Agile}      & \textbf{Agile} & \textbf{Plan-} & \textbf{Plan-} & \textbf{Plan-}\\
                       &                   &  \                    &                           &                       & \textbf{driven} & \textbf{driven} & \textbf{driven}\\
		      & \textbf{All}  & \textbf{Small} & \textbf{Medium} & \textbf{Large} & \textbf{Small}         & \textbf{Medium} & \textbf{Large}\\ 
\textbf{Cause}& \textbf{228} & \textbf{30}      & \textbf{22}         & \textbf{39}       & \textbf{11}             & \textbf{4} & \textbf{30}\\ \hline

Lack of time & 42 (18\%) & 2 (7\%) & \textbf{6 (27\%)}  & 3 (8\%) & \textbf{3 (27\%)} & 0 (0\%) & \textbf{10 (33\%)} \\
Lack of experience of RE team members & 41 (18\%) & 5 (17\%) & \textbf{6 (27\%)} & 4 (10\%) & \textbf{4 (36\%)} & 0 (0\%) & \textbf{8 (27\%)} \\
Weak qualification of RE team members & 31 (14\%) & 1 (3\%) & \textbf{8 (37\%)} & 1 (3\%) & 2 (18\%) & \textbf{2 (50\%)} & 4 (13\%) \\
Communication flaws between project team and the customer & 30 (13\%) & 2 (7\%) & 2 (9\%) & 6 (15\%) & \textbf{3 (27\%)} & 0 (0\%) & \textbf{7 (23\%)} \\
Requirements remain too abstract & 29 (13\%) & 4 (13\%) & 2 (9\%) & 5 (13\%) & 0 (0\%) & 0 (0\%) & \textbf{6 (20\%)} \\
Changing business needs & 21 (9\%) & 1 (3\%) & 2 (9\%) & 3 (8\%) & 1 (9\%) & 0 (0\%) & 0 (0\%) \\
Customer does not know what he wants & 20 (9\%) & 3 (10\%) & 0 (0\%) & \textbf{8 (21\%)} & 0 (0\%) & \textbf{1 (25\%)} & 1 (3\%) \\
Missing direct communication to customer & 18 (8\%) & 1 (3\%) & 4 (18\%) & 3 (8\%) & 1 (9\%) & 0 (0\%) & 0 (0\%) \\
Language barriers & 17 (7\%) & 0 (0\%) & 1 (5\%) & 3 (8\%) & 1 (9\%) & 0 (0\%) & 4 (13\%) \\
Strict time schedule by customer & 16 (7\%) & 2 (7\%) & 0 (0\%) & 4 (10\%) & 0 (0\%) & 0 (0\%) & 2 (7\%) \\
\bottomrule
\end{tabular}
\end{table}

We can observe that the main reported causes of RE problems are \emph{Lack of time}, \emph{Lack of experience of RE team members}, and \emph{Weak qualification of RE team members}. While none of the causes was cited by more than 20\% of the organisations, this figure changes within the specific clusters, where some causes were commonly reported. The cause frequencies above 20\% within each cluster are highlighted in bold.  What can also be seen, even if implicitly, are cycles in the causes and the problems, i.e. some of the causes are, in fact, problems; for instance, \emph{Communications between project team and the customer} is given as one problem, but also named by our respondents as a cause. 

\begin{figure}[!hbt]
\centering
  \includegraphics[width=1\textwidth]{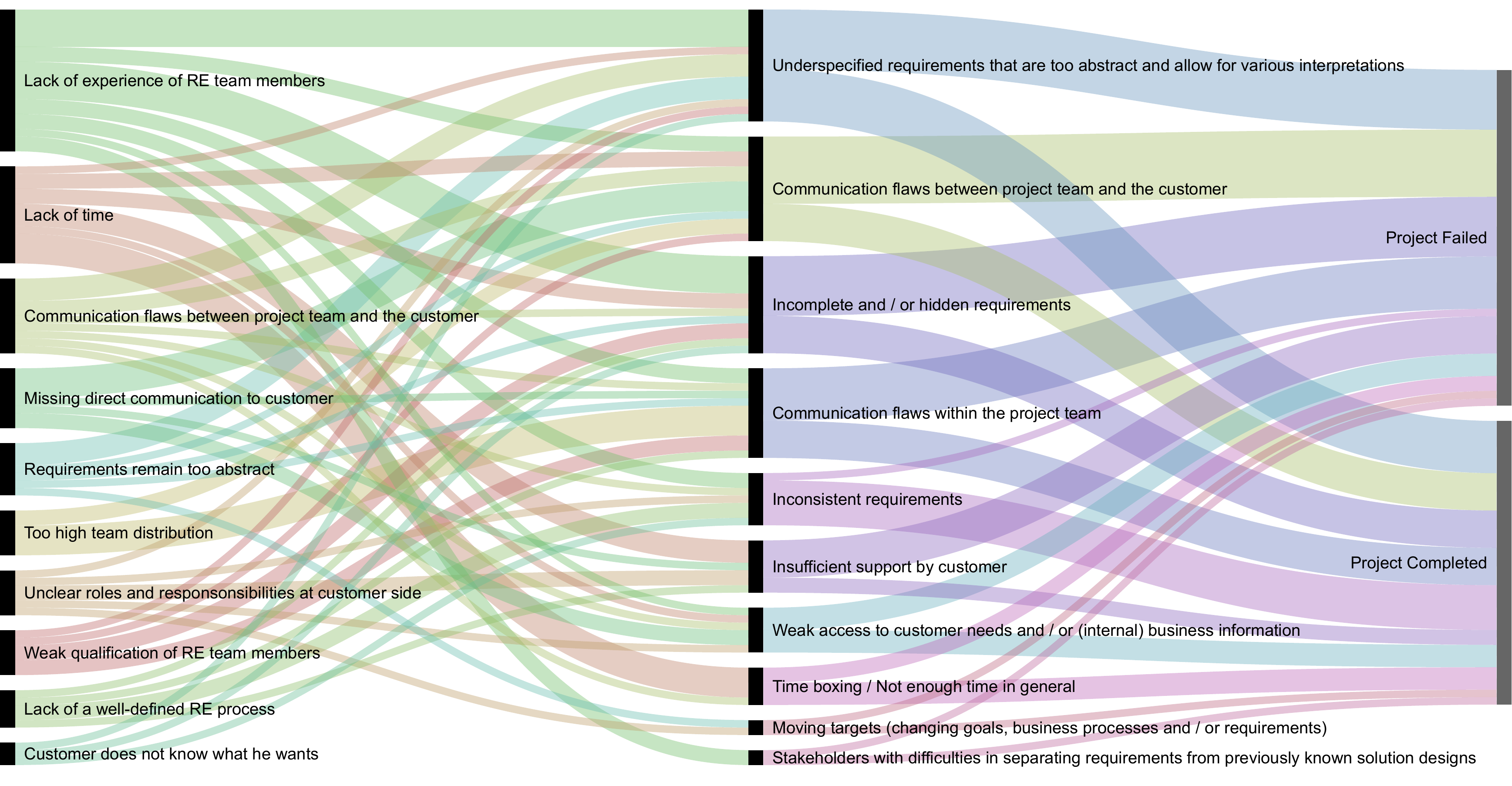}\\
  \caption{Relation of top 10 causes, top 10 problems, and the project impact.}\label{fig.causes_impact_alluvial}
\end{figure}

To analyse the influence of the most cited causes on the most cited problems and, in turn, of those problems to project failure (as reported by the survey respondents), we visualise the relationships via an alluvial diagram. This diagram is shown in Figure~\ref{fig.causes_impact_alluvial}. The decision to relate only the most cited causes to the most cited RE problems was taken to enhance the visualisation.

As it appears, some of the ten most cited causes are more related to some specific problems than to others. Typical examples that can be seen are \emph{Lack of time} leading mainly to the \emph{Time boxing problem}, \emph{Lack of experience of RE team members} leading mainly to \emph{Incomplete and / or hidden requirements} and \emph{Underspecified requirements}, or \emph{Too high team distribution} leading mainly to \emph{Communication flaws within the project team}.

Concerning project failure, the occurrence of some RE problems seems to lead very often to project failure. Out of these, we highlight \emph{Communication flaws between project team and the customer}, \emph{Incomplete and / or hidden requirements}, \emph{Underspecified requirements}, \emph{Communication flaws within the project team}, and \emph{Insufficient support from the customer}. However, in particular relating to project failure, it is noteworthy that this diagram is based on a reduced dataset, considering only instances which contain one of the main causes, one of the main problems, and a project impact. The complete data on how often each of the most cited problems was related by the respondents to project failure is provided in Table~\ref{tab:mostcriticalreproblems}.

\subsubsection{Causes and Effects of Top RE Problems}
\label{sec:causes-effects-top-problems}

To provide a more complete view on the causes and effects reported for some of the most critical RE problems, in particular, the three most cited ones (which are also the three most cited ones for project failure), \emph{Incomplete and / or hidden requirements}, \emph{Communication flaws between us and the customer}, and \emph{Moving Targets}, we built probabilistic cause-effect diagrams. Those diagrams have already been applied in previous efforts in the NaPiRE context, based on data from Brazil (see also Section~\ref{sec:NaPiREInitiative}). It is noteworthy to mention that, despite of the name of those diagrams, within this paper we use them to represent relative frequencies, i.e. how often each cause or effect was cited out of the total citations, and not probabilities. 

\begin{figure}[!hbt]
\centering
  \includegraphics[width=1\textwidth]{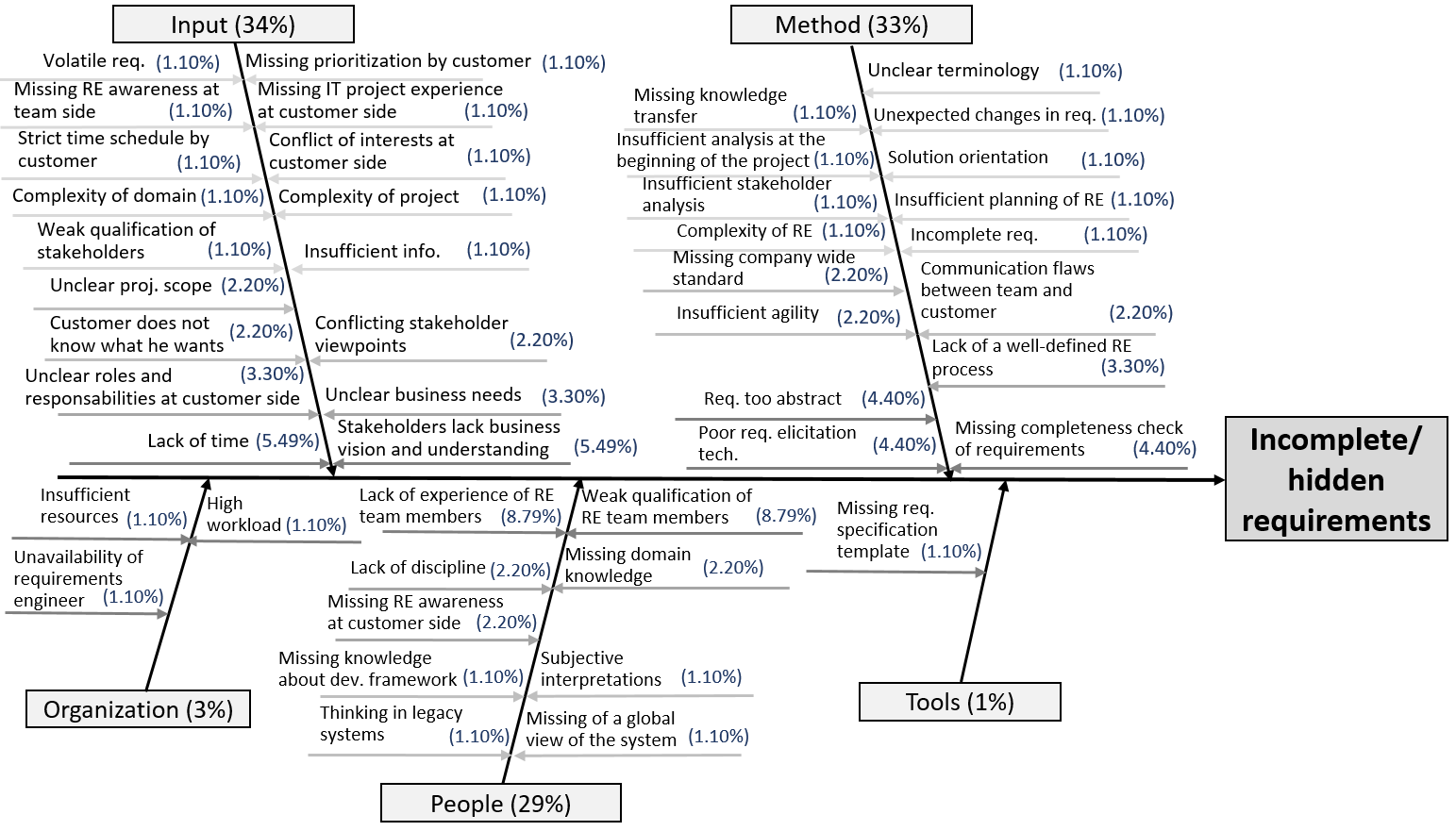}\\
  \caption{Probabilistic cause-effect diagram for \emph{Incomplete and / or hidden requirements} focussing on the causes.}\label{fig.causes_incompletereqs}
\end{figure}

\begin{figure}[!hbt]
\centering
  \includegraphics[width=1\textwidth]{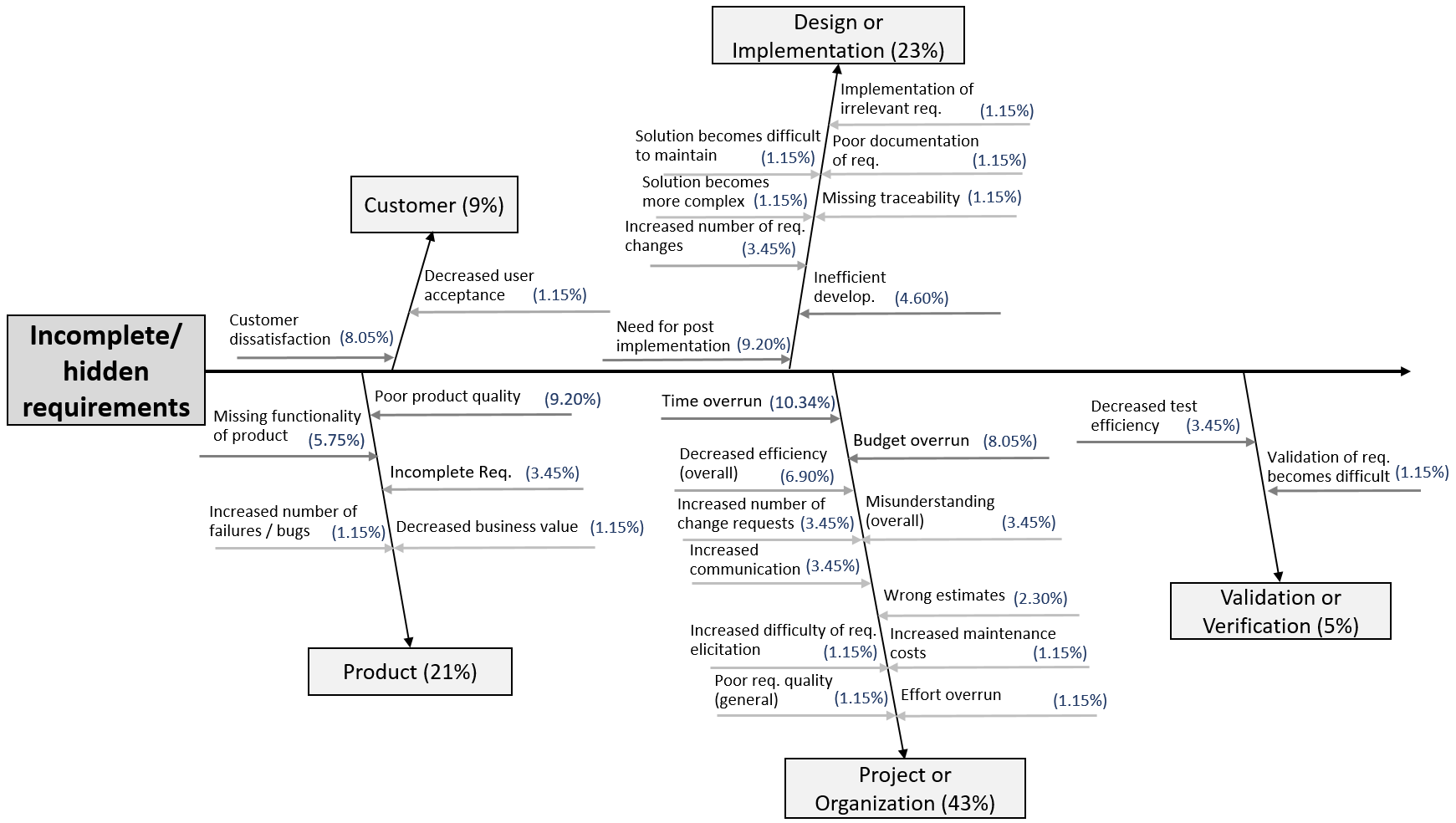}\\
  \caption{Probabilistic cause-effect diagram for \emph{Incomplete and / or hidden requirements} focussing on the effects.}\label{fig.effects_incompletereqs}
\end{figure}

Figures~\ref{fig.causes_incompletereqs} and \ref{fig.effects_incompletereqs} respectively show such cause-effect diagrams for the causes and the effects of the \emph{Incomplete and / or hidden requirements} problem. For instance, in Figure~\ref{fig.causes_incompletereqs}, we can see that the most frequently cited causes were related to the categories \emph{Input} ($\sim34$\%, i.e. 31 out of 91 reported causes were from that category), \emph{Method} ($\sim33$\%), and \emph{People} ($\sim29$\%). The five most frequent reported causes for this problem are the \emph{Weak qualification} ($\sim9$\%) and the \emph{Lack of experience} ($\sim9$\%) of the RE team members, \emph{Time pressure} ($\sim5$\%), \emph{Stakeholders lacking business vision and understanding} ($\sim4$\%), the use of \emph{Poor requirements elicitation techniques} ($\sim4$\%), \emph{Specifying the requirements in an too abstract way} ($\sim4$\%), and \emph{Missing completeness checks} ($\sim4$\%).

Concerning the effects of this problem, it can be seen in Figure~\ref{fig.effects_incompletereqs} that the main affected categories were \emph{Project or Organization} ($\sim43$\%, i.e. 37 out of 87 reported effects were from that category), \emph{Design or Implementation} ($\sim23$\%), and \emph{Product} ($\sim21$\%). The most frequently cited causes were \emph{Time overrun} ($\sim10$\%), \emph{Post implementation rework} ($\sim9$\%) and \emph{Poor product quality} ($\sim9$\%). 

\begin{figure}[!hbt]
\centering
  \includegraphics[width=1\textwidth]{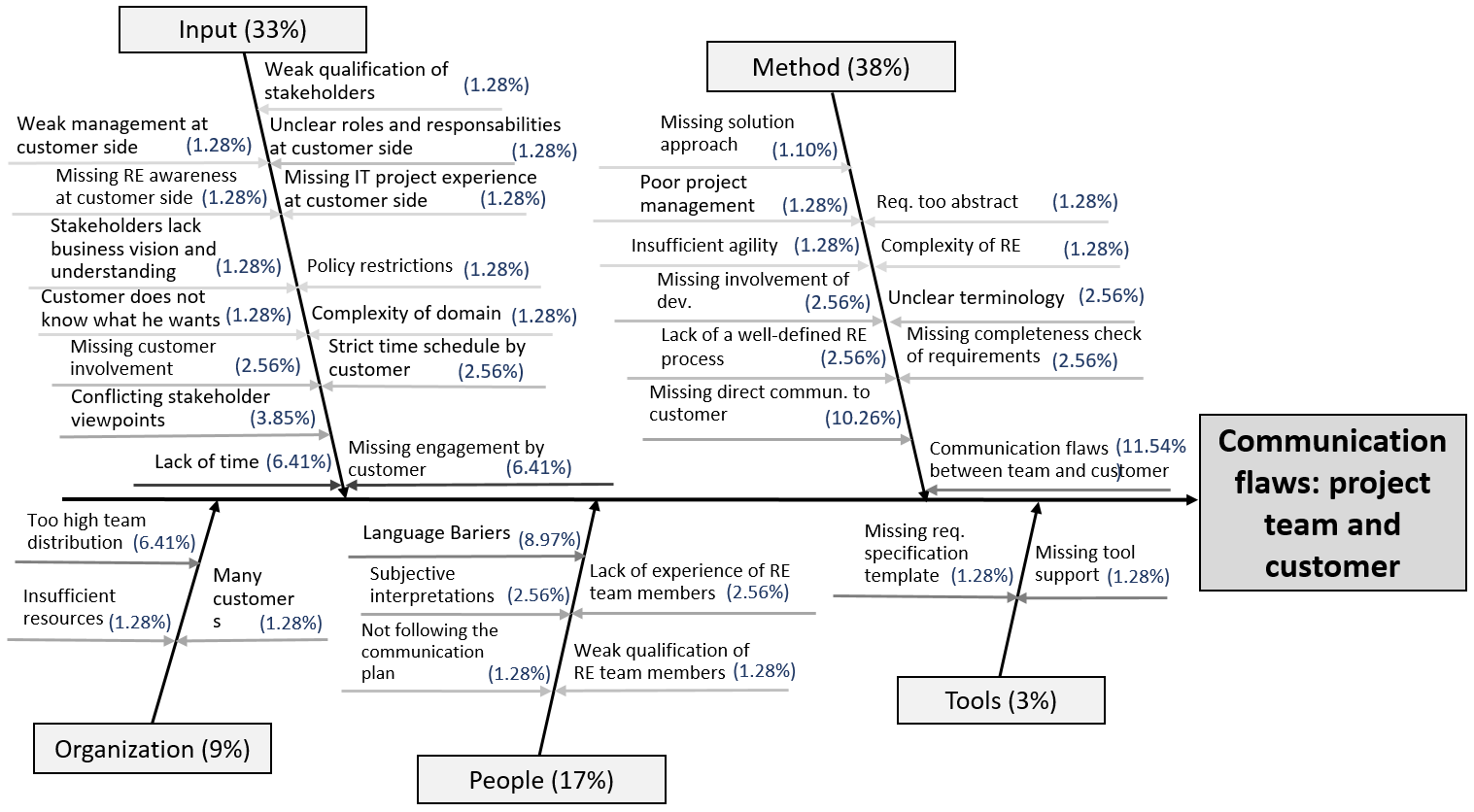}\\
  \caption{Probabilistic cause-effect diagram for \emph{Communication flaws between project team and customer} focussing on the causes.}\label{fig.causes_CFPC}
\end{figure}

\begin{figure}[!hbt]
\centering
  \includegraphics[width=1\textwidth]{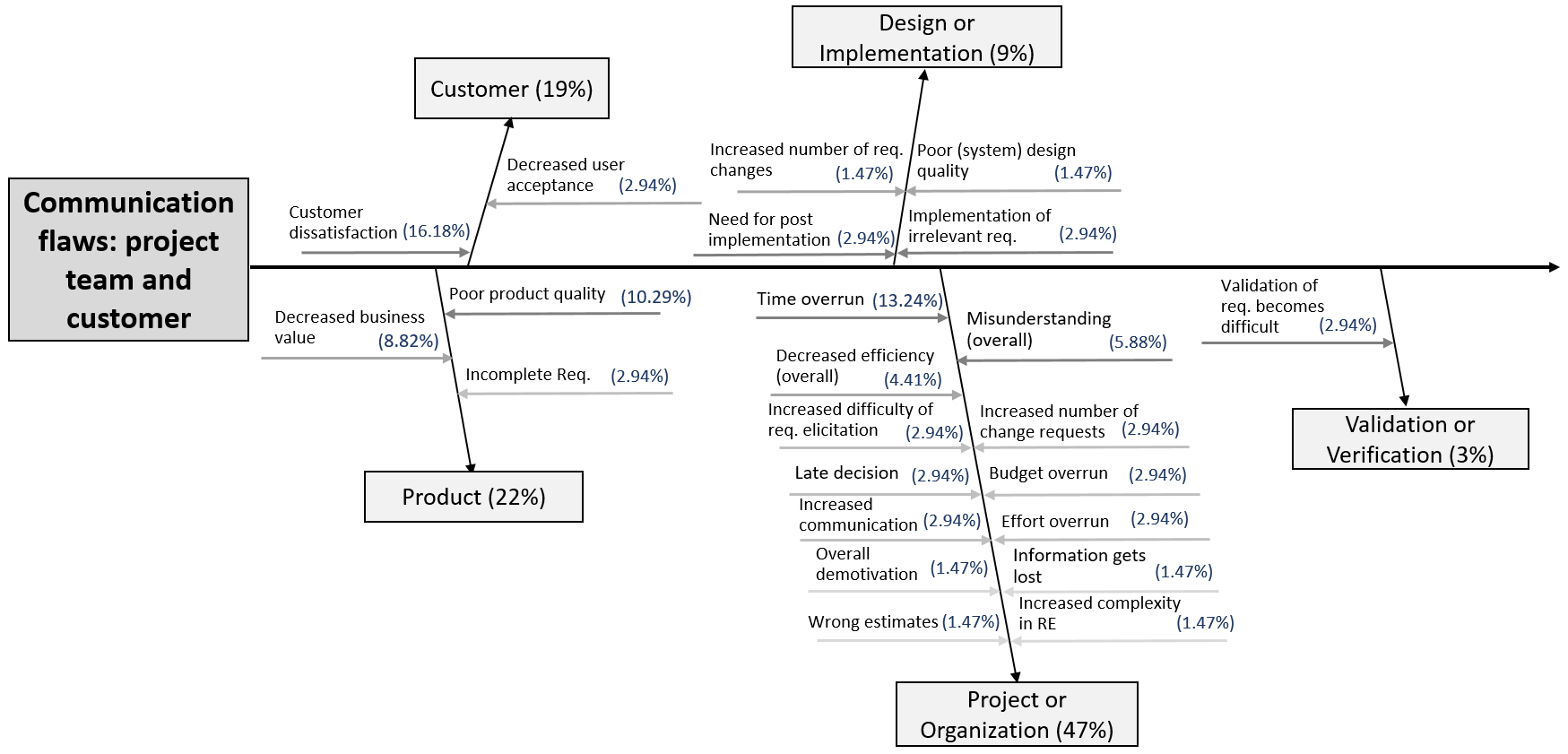}\\
  \caption{Probabilistic cause-effect diagram for \emph{Communication flaws between project team and customer} focussing on the effects.}\label{fig.effects_CFPC}
\end{figure}

The causes and effects of the \emph{Communication flaws between project team and the customer} problem are depicted in Figures~\ref{fig.causes_CFPC} and \ref{fig.effects_CFPC}. The prevailing cause categories for this RE problem are \emph{Method} ($\sim38$\%, i.e.\ 30 out of 78 cited causes were from that category) and \emph{Input} ($\sim33$\%). The five most frequently reported causes for this problem are \emph{Inherent communication flaws} ($\sim12$\%), \emph{Missing direct communication} ($\sim10$\%), \emph{Language barriers} ($\sim9$\%), \emph{Time pressure} ($\sim6$\%), \emph{Missing engagement by the customer} ($\sim6$\%), and a \emph{Too high team distribution} ($\sim6$\%).

In this case (Figure~\ref{fig.effects_CFPC}), the main affected categories were \emph{Project or Organization} ($\sim47$\%, i.e. 32 out of 68 effects were from that category), \emph{Product} ($\sim22$\%), and \emph{Customer} ($\sim19$\%). The main cited effects for this problem were \emph{Customer dissatisfaction} ($\sim16$\%), \emph{Time overrun} ($\sim13$\%), and \emph{Poor product quality} ($\sim10$\%). 

\begin{figure}[!hbt]
\centering
  \includegraphics[width=1\textwidth]{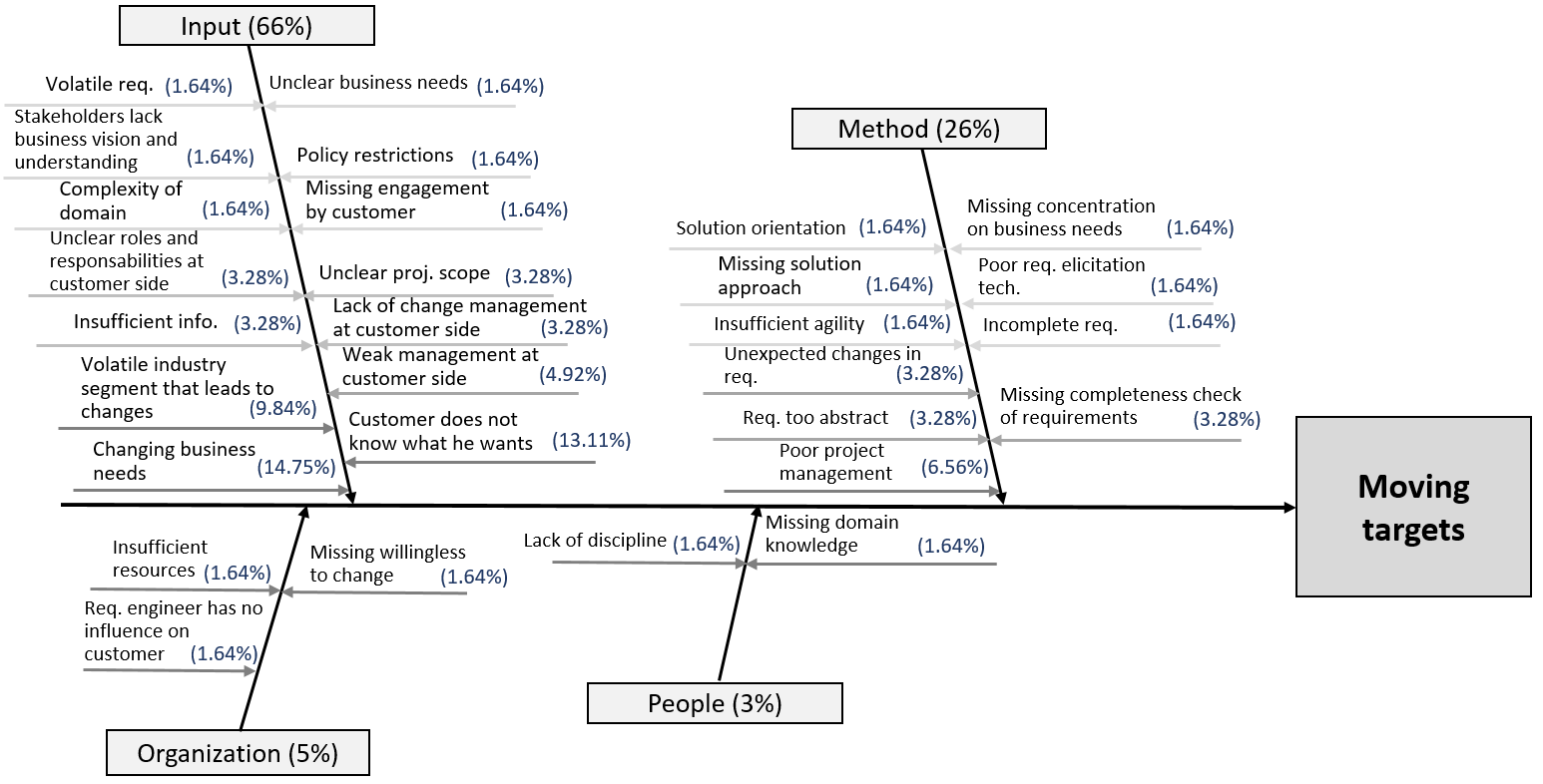}\\
  \caption{Probabilistic cause-effect diagram for \emph{Moving targets} focussing on the causes.}\label{fig.causes_MT}
\end{figure}

\begin{figure}[!hbt]
\centering
  \includegraphics[width=1\textwidth]{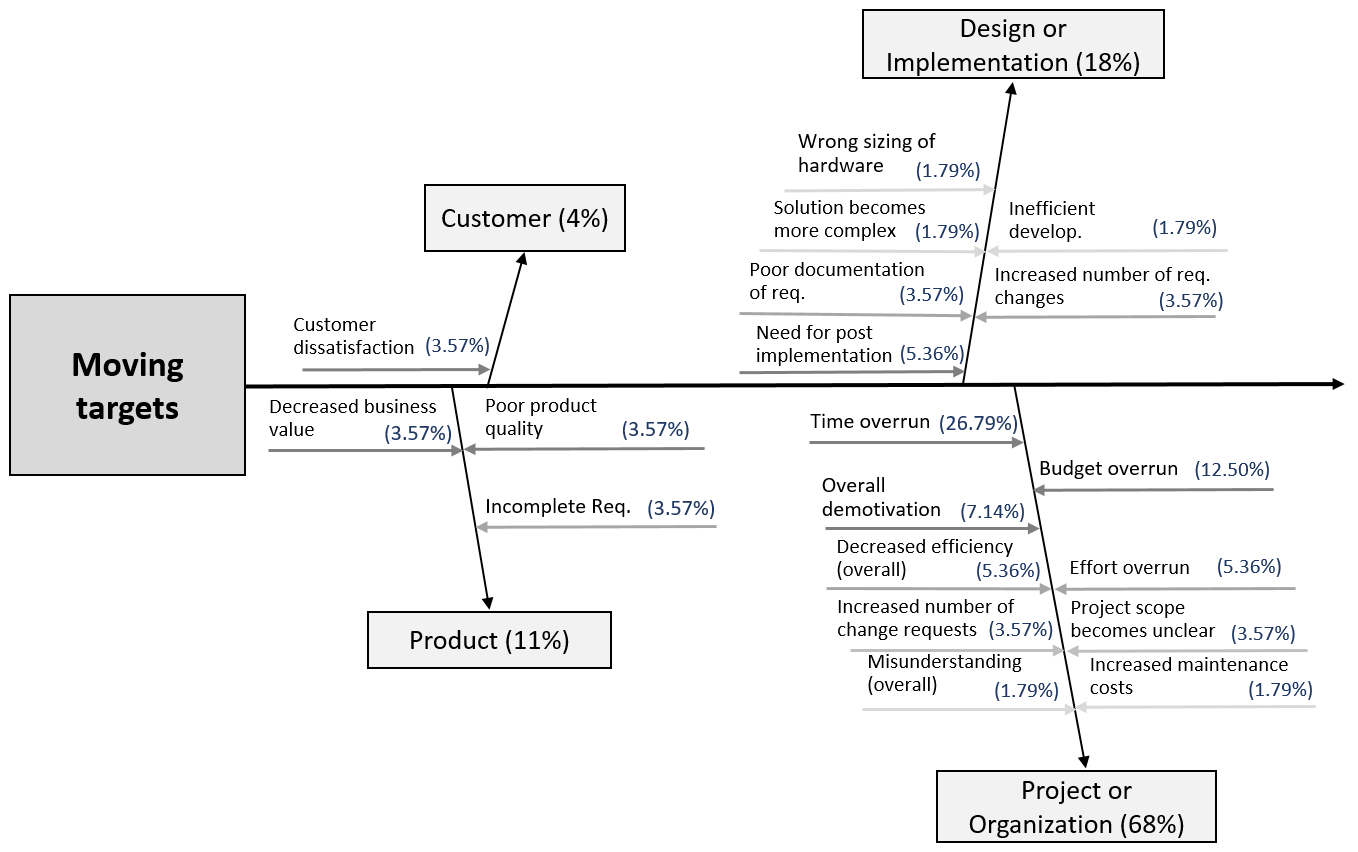}\\
  \caption{Probabilistic cause-effect diagram for \emph{Moving targets} focussing on the effects.}\label{fig.effects_MT}
\end{figure}

Finally, Figures~\ref{fig.causes_MT} and \ref{fig.effects_MT} show the probabilistic cause-effect diagram for the causes and the effects of \emph{Moving Targets}. As shown in Figure~\ref{fig.causes_MT}, the causes of this problem are heavily concentrated on the \emph{Input} category ($\sim66$\%, i.e. 40 out of 61 cited causes were from that category). Its main causes were coded to \emph{Changing business needs} ($\sim15$\%), \emph{Customers who do not know what they want} ($\sim13$\%), \emph{Volatile industry segments that lead to changes} ($\sim10$\%), \emph{Poor project management} ($\sim7$\%), and \emph{Weak management at the customer side} ($\sim5$\%).

Again, as expected for one of the most cited RE problems, the effects are severe (Figure~\ref{fig.effects_MT}). Most of the effects are concentrated in the \emph{Project or Organization} ($\sim68$\%, i.e. 38 out of 56 cited effects were from that category) category, which might explain why this problem has such a strong relation to project failure. In fact, 51\% of the organisations that cited \emph{Moving Targets} as a problem stated that it led to project failure. In this case, the effect \emph{Time overrun} is clearly the most cited one ($\sim27$\%), followed by \emph{Budget overrun} ($\sim13$\%), and \emph{Overall demotivation} ($\sim7$\%). 

\section{Conclusions}
\label{sec:Conclusions}

In this article, we contributed the analysis of contemporary problems practitioners experience in their industrial project setting. To this end, we relied on the NaPiRE initiative (\url{http://www.re-survey.org}), a global family of bi-yearly replicated surveys where the aim is to overcome the problem of by now isolated investigations in RE that are not yet representative. Our analysis of contemporary problems uses data provided by 228 companies spread over 10 countries and included an investigation of problems practitioners experience, what the causes of those problems are, and how the problems manifest themselves in the process going beyond simplified views on project failure.

\subsection{Discussion of Results}

In this section, we discuss our research questions based on the obtained survey results and their potential implications.

\paragraph{Problems in RE (RQ~1)}
Our first research question concerned the contemporary problems in RE. We identified and ranked the problems cited as the most critical ones by the 228 organisations that completed the survey. This result reflects the contemporary opinion of organisations spread throughout ten different countries, of different sizes and using different process models. We believe that this result provides further insights into industrial RE problem trends and that it helps to lay the foundation to steer academic and industrial research in a problem-driven manner where scientific contributions to RE can be put in tune with practically relevant problems. Out of the identified problems, we highlighted \emph{Incomplete and / or hidden requirements}, \emph{Communication flaws between project team and the customer}, and \emph{Moving targets}, which were the most cited ones and also the ones mostly related to project failure.

\paragraph{Problem Patterns (RQ~2)}
The second research question relates to identifying patterns between problems and context characteristics. Of course, there are several ways of blocking our results that could have been performed for this analysis. In this initial effort, we focused on the type of process model (agile or plan-driven) and on the organisational size, blocking those clusters of survey responses. Within these blocks, it was already possible to observe some relevant behavioural differences. For instance, considering the three most cited problems, \emph{Incomplete and/or hidden requirements} appears in all clusters, while \emph{Communication flaws between project team and the customer} does not seem to be a major problem for small plan-driven organisations and \emph{Moving targets} occurs mainly in plan-driven and in large organisations. Future work includes investing additional effort relating the problems to other relevant constructs of the underlying NaPiRE theory. Based on our initial observations, we believe that specific advice to organisations with different characteristics on how to prevent critical RE problems could be valuable beyond the type of advice available in current guidelines and maturity models.

\paragraph{Cause-Effect Analysis (RQ~3)}
The third research question concerned the causes and the effects as they have been perceived by our respondents. We identified the most reported causes and analysed their influence on the most critical RE problems. We could observe that the causes tend to differ within the selected blocks and that some of the ten most cited causes have more influence on some specific problems than on others. 

Additionally, and still in the context of RQ~3, we analysed the causes and effects related to the three most critical RE problems. We believe that the identification of the causes can already help organisation to focus their prevention efforts. For instance, we identified that, in general, the main reported causes for \emph{Incomplete and/or hidden requirements} are the \emph{Weak qualification} and the \emph{Lack of experience} of the RE team members, \emph{Time pressure}, \emph{Stakeholders without business vision}, \emph{Poor elicitation techniques}, \emph{Too abstract specifications}, and \emph{Missing completeness checks}. Based on this information, an organisation facing this or similar problems could take first counter measures, such as: 
\begin{compactenum}
\item Checking on the qualification and experience of its team members, providing training if needed, in particular, on avoiding abstract specifications. This could also be supported by including and training RE standards that put emphasis on the way requirements should be elicited and specified.
\item Adjusting its portfolio management to avoid accepting projects under extreme time pressure or involving stakeholders that lack business vision.
\item Assessing and improving the efficiency of their elicitation techniques.
\item Improving their completeness check, within the philosophy of their development paradigm. Plan-driven organisations, for instance, could institutionalise requirements inspections based on RE standards for the artefacts, while agile organisations could introduce the Definition of Ready (DoR) practice, which is commonly used in agile projects to avoid the beginning of work on features that do not comply with clearly defined completion criteria.
\end{compactenum}

Future work in this direction includes setting up a knowledge base on typical causes of RE problems and on actions that could be taken to mitigate or prevent them (success factors). However, we reinforce that these are informal suggestions of the authors based on the identified causes and that our analyses need to be backed up by complementary investigations, ideally also by applying different empirical research methods on project data (e.g., case studies and experiments). Moreover, organisations should perform in depth causal analysis in their projects to assure addressing the right causes, the ones that are really happening in their concrete context.

\subsection{Relation to Existing Evidence}

In this section, we relate the results of this article to evidence from previous NaPiRE studies and other related RE surveys presented in Section~\ref{sec:RelWork}.  

\paragraph{Previous NaPiRE Evidence}

The first NaPiRE run with data from 58 respondents from Germany \cite{MW14} provided a very similar picture of the top problems as we found
here. The by far most cited problem in both was surveys \emph{incomplete / hidden requirements}. \emph{Moving targets}, 
\emph{time boxing} and \emph{underspecified requirements are too abstract} being in the top 5 in both surveys. \emph{Difficulty of
separating requirements from known solutions}, \emph{inconsistent requirements} and \emph{communication flaws in the team}
occurred in both top tens with a slightly different ranking. Most interesting is \emph{communications flaws between team and customer}.
It moved from rank 6 in the first run to rank 2 in second run. Furthermore, the first run included \emph{missing traceability} and \emph{gold
plating} (ranks 9 and 10) which were replaced in the second run by \emph{insufficient support by customer} and \emph{weak access to customer
needs} (ranks 8 and 10). Hence, the extension to other countries and a larger sample emphasised mainly customer-related problems.
This suggests that these problems are not so prevalent in Germany. Analysing the data from the second run in this respect somewhat
supports this: The problems \emph{insufficient support by customer} and \emph{weak access to customer needs} are mentioned less often than in
most other countries. A reason could be that more often than in other countries, the team and the customer are geographically close and
speak the same language. Yet, \emph{communication flaws between team and customer} are also very prevalent in the German data of
the second run.

In \cite{kalinowski:seke15}, the the Brazilian data set was analysed w.r.t.\ the cited problems, causes and effects. The set of top problems in Brazil
matches exactly the top problems in the global data set. The ranking of the problems differs only slightly. There are differences in the causes
and effects, but the general categories are consistent with some variation in the weights.

The data sets from Brazil and Germany were also compared separately~\cite{mendez:softw15}. The top five problems from both countries
are in the top 8 of the global data set. Also all other results are very similar to the results from the full data set. 

Finally, the data sets from Brazil and Austria was used to investigate in detail the \emph{incomplete / hidden requirements} 
problem~\cite{kalinowski:swqd16}. The causes found there are also included in the results of this article. The distribution over the categories 
of causes changed slightly from a strong focus on people (40\%) to input (34\%) and method (33\%). Yet, the results of this article contain 
far more causes than we found in~\cite{kalinowski:swqd16}.

\paragraph{Further Existing Evidence}
Our results corroborate and extend existing findings reported by other researchers. For example, the German Success~study~\cite{success07} investigated factors that influence project success in general, not limited to requirements
engineering. Most of the factors are consequently more abstract than ours. However, the study found that project success is independent of the degree of management support. We can support this to a certain extent in the sense that missing management support was not often perceived as a problem.

Kamata et al.~\cite{IT07} investigate relations between requirements quality and project success or failure, which aimed at a more limited view than we took as the requirements specification was in scope of their investigation rather then the whole RE process. Therefore, the results are not directly comparable. Nevertheless, their finding that a relatively small set of requirements has strong impact on project success or failure is relevant for our top RE problem of incomplete and / or hidden requirements, because it further substantiates the problem of single missing and / or hidden requirements.

Nikula et al.~\cite{nikula2000sps} find in their state of practice survey on requirements engineering in small- and medium-sized companies that completeness, change management and descriptions are the three most needed techniques to further develop RE in the participating companies. These three top needs are directly related to our three top RE problems and, thus, further underpin their high rating in our survey and the relevance of the provided cause-effect analysis for these problems in Section~\ref{sec:causes-effects-top-problems}. 


Solemon et al.~\cite{solemon2009requirements} report requirements engineering problems and practices in software companies. Differing from our study, the focus of the researchers was to identify RE problems rather than relating them to possible causes. They classified RE problems into organisational and RE process related ones. The reported main organisational problems are lack of customer and user communication problem, lack of developer communication, as well as poor time and resources allocation. The main problems in the RE process are related to changing requirements, incomplete requirements, ambiguous requirements and poor user understanding. With regard to the mentioned communication, resource, change, completeness and understandability issues, our survey supports their findings, and additionally provide a more fine-grained distinction and relationships between causes and effects.

Liu et al.~\cite{liu2010requirements} present results of a survey on why requirements engineering fails. According to that survey, major failure reasons are an unclear understanding of the system by the customer, constant change of user needs and understanding, missing access to domain knowledge for software engineers, reuse of existing design in wrong context and environment, lack of domain and technical expertise for RE decision makers, tight project schedule, broken communication links, as well as lack of standardised data and interface definitions. The authors do not clearly distinguish between problems and their causes, but all listed major failure reasons are related to our main RE problems and their causes.

Verner et al.~\cite{verner2007requirements} further ran a survey in Australia and the USA. They concentrated on success factors
in RE and found good requirements, customer/user involvement, and effective requirements management to be
the best predictors of project success. Disregarding the difficulties to precisely capture what ``good requirements'' are, our results still can be considered in tune with their observations as we identified problems and their causes which (if negated) can be used to refine the abstract success factors identified by Verner et al. For instance, we identified incomplete and/or hidden requirements as a main problem and weak qualification as well as lack of experience as its main causes. 

Al-Rawas and Easterbrook \cite{al1996communication} finally present a field study on communication problems in requirements engineering. The results show that organisational issues have great influence on the effectiveness of communication, and furthermore that in general users find the notations used by software practitioners to model their requirements difficult to understand and validate. The topic of the study shows that communication is an important RE problem, which is also reflected by our results. Also the presented results are in tune with the causes for communication flaws that we present.


\subsection{Impact/Implications}

Our findings complement existing evidence on problems in RE in various ways. First, we could distill a detailed picture of problems practitioners experience in their project setting including a rich analysis of effects going beyond project failure. Second, the analysis of the causes and effects in dependency to context factors allows us to steer a first empirically founded discussion on phenomena that hold for particular contexts. Third, and most importantly, revealing not only the problems, but also their causes, allows us to get a first picture of requirements engineering success factors which, if met, should mitigate the problems. 

Based on this analysis, we can already steer further problem-driven research, for example on agile requirements engineering. It also allows us to further explore RE success factors suitable to establish maturity models and RE improvement endeavours that are grounded on empirical data. 

Overall, our contribution does not only support researchers to steer their research, but also practitioners to evaluate their own current RE situation against overall industrial trends presented here.

\subsection{Limitations}

We have analysed the results from the survey conducted in ten countries yielding a broad population, while applying the different procedures to control the validity as described in section~\ref{sec:ValidityProcedures}. Still, we are aware that our study has limitations. Most importantly, our results are based on a reasonable but still limited number of respondents. Also, the responses are not equally distributed over business domains and families of systems (such as embedded systems). This probably has an influence on the rankings of the problems if specific domains have specific problems. We can, for example, not say reliably how the picture would change if considering safety-critical systems only, let alone as they appear less frequently in our data than information systems. We can only assume that the picture would change for practices that tend to be seen as more important for the development of highly dependable, safety-critical systems than for business information systems (such as traceability). Furthermore, we cannot make concrete statements about how generalisable the results eventually are, because we still are not able to estimate the representativeness of our population (given the unavailability of empirical data characterising all industry segments in every considered country). This means in consequence that we cannot generalise going beyond the contexts described and that we might even expect partially different results in different countries. Therefore, we need to follow our design of a family of surveys and further steer the continuous replications and syntheses of the results while capturing precisely the context to establish a more reliable and empirically solid theory. 

Furthermore, inherent to survey research is that surveys can only reveal stakeholders' perceptions on current practices rather than empirically backed-up knowledge about those practices. To some extent, we aim at revealing exactly those perceptions. However, the answers given by our respondents might still be biased. We mitigated this threat by conducting the survey anonymously, but need to back up (and further explore) the insights revealed via this survey by adding more investigations using other empirical methods, e.g., interviews and case studies.

A further limitation arises from the qualitative data and its analysis. Manual coding of qualitative data is by nature a creative process where experiences, expectations and the expertise of the coders influence the results. We mitigated this threat to a certain extent by applying researcher triangulation, yet the picture might still change if involving other researchers. 

Finally, a major threat arises from the way we designed our instrument as it still includes too many (to a certain extent closed) questions which might enforce too simplified views by the respondents on the particularities of their project environments. We asked our respondents to categorise their software process model used based on a pre-defined list of options. This means that although the respondents might have selected, for instance, Scrum, we still have no guarantee that the process model followed was indeed Scrum or even agile. It could have also been a plan-driven, iterative model baptised as ``Scrum'' in that organisational context. That is, respondents might have misinterpreted the options or they might have relied on a different understanding on the process model used (or even on how the process model should be rather than how it actually is), let alone because of the fuzzy notion of ``agile'' that allows for too many variations. So far, we cannot mitigate the threats arising from this circumstance, but use the lessons we learnt in this replication to further foster our learning curve during the follow-up re-redesign of our instrument. Similarly, the part of our instrument used to reveal problems relies, from a conceptual point of view, on a simplified understanding of causes and effects. We realised that such a simplified view on phenomena and their sequential cause-effect dependencies is not sufficient to capture the complex dependencies of the various factors in a project; we need, in fact, to extend our instrument with means that allow to capture complex systems of phenomena as recommended in context of system theory (and system thinking respectively). The re-design of our instrument therefore includes, inter alia, a more indirect approach that relies on triangulation from multiple questions to support richer and more robust results (see also our next section).

\subsection{Future Work}
\label{sec:FutureWork}
We are currently analysing the overall data obtained from the current NaPiRE survey round with respect to the responses on the status quo in RE. This current analysis complements the investigation presented here with a first holistic theory on the current status quo in RE. Another work already in progress is the usage of results from the cause-effect analysis to establish a maturity model for RE including practical recommendations on how to improve RE in response to certain context-specific situations. 

Besides current analyses of the survey data and subsequent work based on the data, we are planning the next survey round to be started in 2017. To this end, we will first revise our instrument to better mitigate the limitations discussed above. This re-design includes:
\begin{compactitem}
\item A richer set of questions to describe the context of our respondents including a more specific focus on teams and project environments rather the overall organisation.
\item A more differentiated view on the particularities of the projects and the process models used including a more indirect approach that relies on triangulation from multiple, more objective questions on the practices actually used. 
\item A more differentiated view relying on more open questions following an approach as recommended by the 5-why methodology to capture the complex system of dependent phenomena rather than relying of single causes and effects of a set of pre-compiled problems.
\end{compactitem}

Given that NaPiRE is a community effort aimed at bringing forward the empirical RE community, we cordially invite researchers and practitioners to join us in our follow-up activities to steer future replications of our family of surveys.

\subsection{Acknowledgements}
We want to thank all members of ISERN who supported the design of this research initiative. Furthermore, we want to thank all our contacts from industry for their participation in the survey. Our gratitude also goes to the reviewers whose constructive feedback supported us improving our manuscript and in particular the further re-design of our instrument for the follow-up replications. Finally, Dietmar Pfahl was supported by the Estonian Research Council.

\bibliographystyle{spbasic}
\bibliography{Literature}

\appendix
\section{List of Contributors and Roles}

Table~\ref{tab:authorshipdetails} introduces the details of the authorships with respect to the roles taken in the NaPiRE project and in context of this particular manuscript. To this end, we introduce a role concept based on the classification scheme as discussed by Brand et al. in~\cite{Brand15}. In our context, we distinguish the following roles\footnote{Those roles marked with an $*$ are the exact same as introduced in~\cite{Brand15}.}:
\begin{compactitem}
\item Conceptualisation$^{*}$: Ideas; formulation or evolution of overarching research goals and aims.
\item Project Administration$^{*}$: Management and coordination responsibility for the research activity planning and execution.
\item Methodology$^{*}$: Development or design of methodology; creation of models.
\item Instrument Design: Development / re-design of the instrument used in this replication. 
\item Data Collection: Data collection as national representative in the respective country using the provided infrastructure.
\item Data Analysis: Application of textual analysis techniques to study and interpret the data.
\item Data Curation$^{*}$: Management activities to annotate (produce metadata), scrub data and maintain research data (including software code, where it is necessary for interpreting the data itself) for initial use and later reuse.
\item Data Visualisation$^{*}$: Preparation, creation and/or presentation of the published work, specifically visualisation/ data presentation.
\item Writing - Original Draft$^{*}$: Preparation, creation and/or presentation of the published work, specifically writing the initial draft (including substantive translation).
\item Data - Review \& Editing$^{*}$: Preparation, creation and/or presentation of the published work by those from the original research group, specifically critical review, commentary or revision – including pre- or post-publication stages
\end{compactitem}

The first two authors are the initiators and coordinators of the overall initiative. Further, we formed a group of lead authors for this particular manuscript (the first six authors) to do the data analysis and / or the writing process.  

\begin{table}[htb]
\scriptsize
\centering
\caption{Authorship details: Roles and contribution.}
\label{tab:authorshipdetails}
\begin{tabular}{p{0.45\linewidth}cccccccccc}
\toprule
\textbf{Author}  & \rot{\textbf{Conceptualisation}} & \rot{\textbf{Project Administration}} & \rot{\textbf{Methodology}} & \rot{\textbf{Instrument Design}} & \rot{\textbf{Data Collection}} & \rot{\textbf{Data Analysis}} & \rot{\textbf{Data Curation}} & \rot{\textbf{Data Visualisation}} & \rot{\textbf{Writing - Draft}} & \rot{\textbf{Writing - Review \& Editing}}\\ \hline
D. M\'{e}ndez Fern\'{a}ndez & X & X & X & X &X & X & X & X& X & X \\
S. Wagner & X & X & X & X & X & X & &  & X & X\\
M. Kalinowski & & & &X&X & X& X & X &X & X\\
M. Felderer & & & & X& X&X &  & X &X & X\\
P. Mafra & & & & & &X & & & &X\\
A. Vetr\`{o} & & & && &  &X & & & X\\
T. Conte & & & &&X &X & & & &X\\
M.-T. Christiansson  & && & &X & & & & &X\\
D. Greer & & & &X& X& & & &&X\\
C. Lassenius & & & & X& X& & & & &X\\
T. M\"annist\"o & & & & X&X & &&  & &X\\
M. Nayabi & & & & & X& & & & &X\\
M. Oivo & & & &X& X& & & &&X\\
B. Penzenstadler & & & & &X & & & & &X\\
D. Pfahl & & & & X& X& & & &  &X\\
R. Prikladnicki & & & & &X & X & & & &X \\
G. Ruhe & & &  && X& & & & &X\\
A. Schekelmann & & & & &X & & & & &X  \\
S. Sen & & & & & X& & & & &X\\
R. Spinola & & & & &X & X &  & & &X\\
A. Tuzcu & & & && & X& & & &X \\
J. L. de la Vara & & & & &X & & & & &X\\
R. Wieringa & & & &X&X & & & & &X \\
\bottomrule
\end{tabular}
\end{table}

\end{document}